\documentclass[prd,preprint,eqsecnum,nofootinbib,amsmath,amssymb,preprintnumbers,tightenlines,longbibliography]{revtex4-1}

\usepackage{mathrsfs}
\usepackage{hyperref}
\usepackage{graphicx}
\usepackage{bm}

\def\otherf{f_{ra-ar}}
\def\three{3}
\def\two{2}
\def\one{1}
\def\ddiff{\overleftrightarrow{\partial}}

\def\Im{{\rm Im}}

\def\kp{{2\pi T_{\rm eff}}}
\def\pp#1{\frac{\partial}{\partial #1} }
\def\rout{{\rho_{\scriptscriptstyle \rm out}}}
\def\ls2{{\ell_s^2}}
\def\x{{\bm x}}
\def\dd{{\rm d}}
\def\Db{{\mathbb D}}
\def\Dt{{\mathrm D}}

\def\f{{ f}}

\newcommand\bga{\begin{align}}
\newcommand\nda{\end{align}}

\def\Tr{\: {\rm Tr} \:}

\def\x{{\bm x}}
\def\hr{{\rho}}
\def\hv{{v}}

\def\F{{\mathcal F}}

\def\F{{\mathcal F}}
\def\N{{\mathcal N}}

\def\Im{{\rm Im}}

\def\eff{{\rm eff}}

\def\nc{N_{\rm c}}

\def\st{\begin{equation}}
\def\stp{\end{equation}}
\def\bg{\begin{eqnarray}}
\def\nd{\end{eqnarray}}
\def\Eq#1{Eq.~(\ref{#1})}
\def\app#1{Appendix~\ref{#1}}
\def\Fig#1{Fig.~\ref{#1}}
\def\Sect#1{Section~\ref{#1}}
\def\Ref#1{Ref.~\cite{#1}}
\def\m{{ m}}

\def\llangle{\left\langle}
\def\rrangle{\right\rangle}

\def\N{\mathcal{N}}

\def\sign{{\rm sign}}

\def\nott#1{\setbox0=\hbox{$#1$}                
   \dimen0=\wd0                                 
   \setbox1=\hbox{/} \dimen1=\wd1               
   \ifdim\dimen0>\dimen1                        
      \rlap{\hbox to \dimen0{\hfil/\hfil}}      
      #1                                        
   \else                                        
      \rlap{\hbox to \dimen1{\hfil$#1$\hfil}}   
      /                                         
   \fi}                                         %

\advance\parskip 1.9pt
\advance\voffset -0.2in

\def\st{\begin{equation}}
\def\stp{\end{equation}}
\def\bg{\begin{eqnarray}}
\def\nd{\end{eqnarray}}
\def\nc{{\, ,}}

\begin{document}


\title{Fluctuation, dissipation, and thermalization in non-equilibrium ${\rm AdS}_5$ black hole geometries }

\author{Simon Caron-Huot}
\affiliation
    {%
School of Natural Sciences, 
Institute for Advanced Study, 
Princeton, NJ 08540, USA 
    }%
\email{schuot@ias.edu}
\author{Paul M. Chesler}
\affiliation
    {%
Department of Physics, 
MIT, 
Cambridge, MA 02139, USA 
    }%
\email{pchesler@mit.edu}

\author{Derek Teaney}
\affiliation
    {%
    Department of Physics \& Astronomy,
    Stony Brook University,
    Stony Brook, NY 11794, USA
    }%
\email{derek.teaney@stonybrook.edu}

\date{January 2011}

\begin{abstract}

We give a simple recipe for computing dissipation
and fluctuations (commutator and anti-commutator correlation functions)   for non-equilibrium black hole geometries. The recipe
formulates Hawking radiation as an initial value problem, 
and is suitable for numerical work.  We show how to package the fluctuation and
dissipation near the event horizon into correlators on the stretched horizon.
These horizon correlators determine the  bulk and boundary field theory
correlation functions.
In addition,  the horizon correlators are  the components of a horizon
effective action which provides a quantum generalization of the membrane
paradigm.  In equilibrium, the analysis reproduces previous results on the
Brownian motion of a heavy quark.  Out of equilibrium,  Wigner transforms of
commutator and anti-commutator correlation functions obey a
fluctuation-dissipation relation at high frequency. 
\end{abstract}

\maketitle

\section{Motivation}

One of the motivations for studying high temperature gauge theories
at strong coupling is the striking results from the Relativistic Heavy Ion 
Collider and the Large Hadron Collider \cite{Adcox:2004mh,Bellwied:2005kq}. Results on collective flow
and the energy loss of energetic probes (in particular heavy quarks \cite{Adare:2010de}) indicate that 
the nuclear size is sufficiently large that macroscopic quantities
such as temperature, pressure and flow velocity, are useful concepts
when characterizing heavy ion events. A back of the envelope 
calculation \cite{Danielewicz:1984ww} shows that this would not be possible
unless the typical relaxation  time is of order a thermal wavelength $\tau_R \sim
\hbar/T$, placing the QCD plasma in a strong coupling regime.
The AdS/CFT correspondence 
has led to many important insights into the
nature of strongly  coupled plasmas  and energy loss \cite{Schafer:2009dj}.
Of particular relevance
to this work is the computation of the heavy quark drag  and 
diffusion  coefficient in $\N=4$ Super Yang Mills (SYM) at
large $N_c$ and strong coupling \cite{HKKKY,Jorge,Gubser}.  
Indeed heavy quarks in heavy ion collisions exhibit a
strong energy loss and a larger than expected elliptic flow,  which is qualitatively
consistent with  a small diffusion coefficient \cite{Adare:2010de}. The actual interpretation
of the RHIC results is more complicated, since radiative energy loss  
plays a significant (perhaps dominant) role at high momentum   where 
the measurements exist \cite{Djordjevic:2004nq}.  
The current results from the Relativistic Heavy 
Ion Collider on heavy quarks are confusing and not generally understood.

The diffusion of heavy quarks in  AdS/CFT  is also interesting 
from the perspective of black hole physics.   Indeed, the primary goal 
of this paper is to  better understand the thermal properties of black holes
using the diffusion of heavy quarks in $\N=4$ SYM as a constrained
theoretical laboratory.

\begin{figure}
\begin{center}
\begin{minipage}[c]{0.48\textwidth}s
\includegraphics[width=\textwidth]{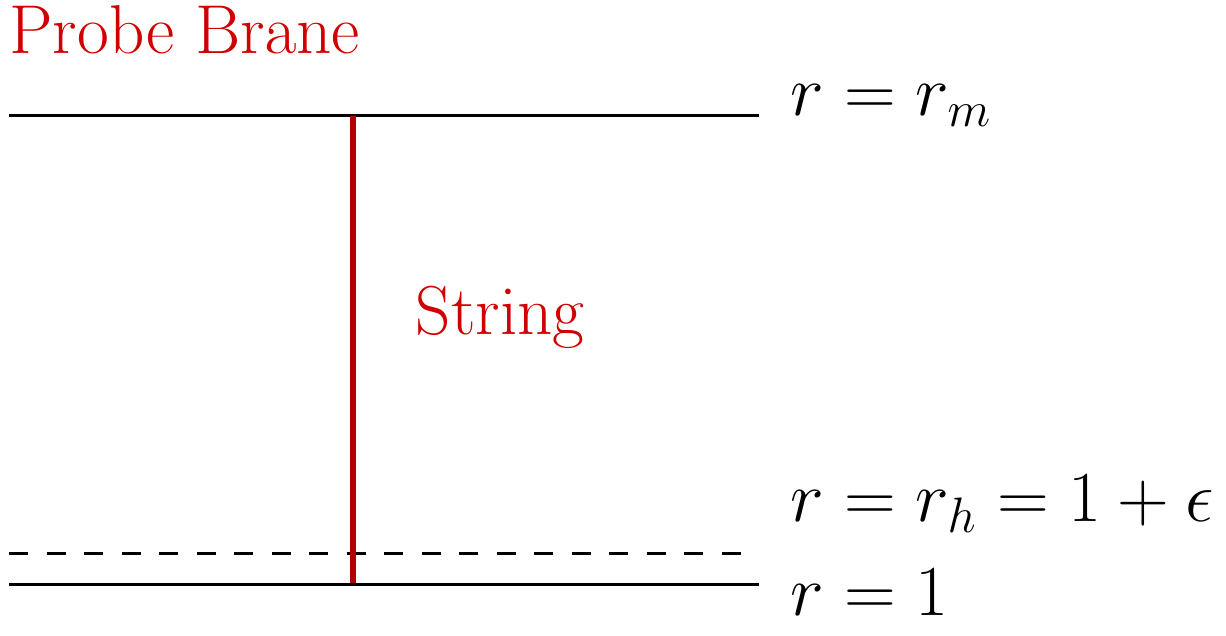}
\end{minipage}
\hfill
\begin{minipage}[c]{0.48\textwidth}
\includegraphics[width=\textwidth]{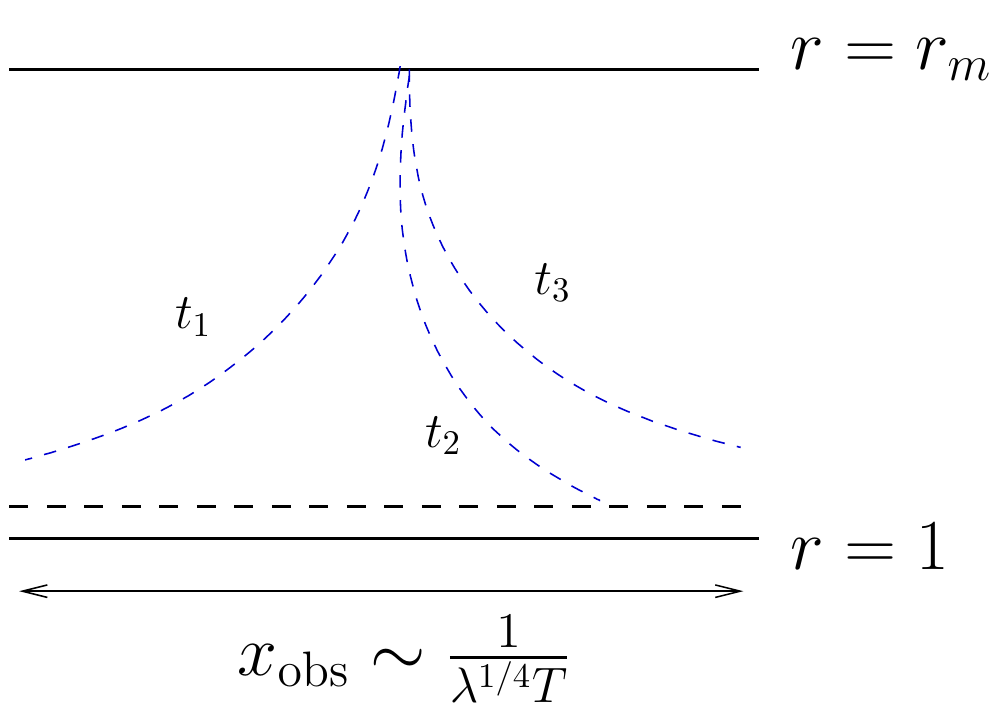}
\end{minipage}
\end{center}
\caption{ \label{sontfig} Two figures from \protect\Ref{Son:2009vu} which motivate this 
work.  
(a) A schematic of a classical string in AdS$_{5}$ corresponding to a
heavy quark.
The horizon is at $r=1$ in the coordinates of this work. The stretched
horizon is at $r_h=1+\epsilon$ and the endpoint of the string is at
the boundary $r_m$ with $r_m \gg 1$. Gravity pulls downward in this figure. 
(b) 
Hawking radiation from the horizon induces stochastic motion
of the string in the bulk which 
we show for three subsequent time steps, $t_1, t_2,t_3$. 
The random string configurations give
 rise to a random force in the boundary theory. 
The Hawking radiation is encoded in an effective action on the 
stretched horizon $r_{h} = 1+ \epsilon$. The string fluctuations
are small, $x_{\rm obs} \sim 1/\lambda^{1/4} T$. } 
\end{figure}

On the field theory 
side of the correspondence, the  diffusion of heavy quarks is the 
result of a competition  between the drag  and the noise, which must 
be precisely balanced if the quark is to reach equilibrium.  In particular, over a time scale which 
is long compared to medium correlations, but short compared to the 
equilibration time of a heavy quark, the heavy quark is expected
to obey a Langevin equation
\st
\label{langevin}
 \frac{dp^{i} }{dt}  =  - \eta v^i  + \xi^i  \, , 
\stp
where the drag coefficient $\eta$ and the random noise 
$\xi$ are balanced by the fluctuation-dissipation relation
\st
\label{einstein}
\qquad \llangle \xi^i(t) \xi^j(t') \rrangle = 2 T \eta \delta^{ij} \delta(t-t') \, .
\stp
Previously it was shown how this Brownian equation of motion 
is reproduced by the correspondence \cite{deBoer:2008gu, Son:2009vu}.
In AdS/CFT
a heavy quark is dual to a long straight string which stretches from the boundary to the  horizon. At a classical level the straight string is a solution 
to the EOM and doesn't move. 
This is not the dual of a heavy quark in plasma.
\Fig{sontfig}(a) shows the geometry of black hole AdS together with
a long straight string. In our AdS conventions the horizon is 
at $r=1$ while the boundary of AdS is at $r=\infty$.  The stretched
horizon (see below) is at $r_h = 1 + \epsilon$.
Hawking radiation from
the horizon causes  the string to flip-flop back and forth 
stochastically as exhibited in \Fig{sontfig}(b).  The random tugs from this flip-flopping string give
rise to a random force  on the boundary quark which is related to
the dissipation  by the Einstein relation, \Eq{einstein}.

The derivation of this result left much to be desired. In \Ref{deBoer:2008gu}
it was simply assumed that the modes are in equilibrium
at the Hawking temperature. With this assumption it is not  difficult
to show that the commutator 
and anti-commutator of string correlations are related via a bulk version
of the Fluctuation Dissipation Theorem (FDT). When this bulk FDT 
is translated to the
boundary theory, the bulk FDT leads to \Eq{einstein}.
While this derivation is physically
reasonable, the calculation provides little guidance to out of equilibrium geometries.
In \Ref{Son:2009vu}, the bulk FDT was derived following
a rather complicated and un-intuitive formalism \cite{Herzog:2002pc,Skenderis:2008dh}. This 
formalism   involves analytically
continuing modes as is typical in many derivations
of Hawking radiation \cite{Hawking:1974sw,Gibbons:1976pt,Gibbons:1976es}.  The primary goal  of this paper is to provide a much more physical derivation
of the bulk FDT based upon propagating near-horizon quantum fluctuations from the distant past up 
to the bulk.

In \Ref{Son:2009vu}  the effects of Hawking radiation were packaged into 
a horizon effective action. This effective action dictates
the dynamics of the fields at $r=1 + \epsilon$, and 
provides a quantum generalization of membrane  paradigm.  Although the derivation
of the membrane effective action involved a complicated analytic continuation, the final form  of the effective action is very natural.  The classical part 
of the effective action can be determined using the classical membrane 
paradigm,  while the quantum part of the effective action is dictated by 
the classical dissipation, and the fluctuation dissipation relation. 
Once the horizon effective action is written down,  a short exercise
shows  how the horizon fluctuation dissipation relation leads
to the FDT in the bulk and boundary theories. 
Since the FDT
is a direct consequence of the fact that the density matrix is $\exp(-\beta H)$, 
the extent to which this relation holds provides an unequivocal measure
of equilibrium in the bulk geometry.  

In this paper we will determine the horizon effective action by solving
equations of motion  with appropriate initial conditions rather
than analytically continuing modes.
Since Fourier transforms and  analytic continuations are never introduced,
it is possible to apply these same techniques to non-equilibrium
geometries.  
We also study the thermalization of a string
in a non-equilibrium geometry.
Furthermore, while we focus 
on string fluctuations in this paper, our analysis easily generalizes to other 
fields such as the graviton.

\section{Preliminaries} 

\label{prelim}

In this subsection we will give a brief summary of  some of the 
results of \Ref{Son:2009vu} in order to establish notation.
The metric of the black hole AdS space is 
\st
ds^2 =  (\pi T)^2 L^2 \left[ -r^2 \f(r) dt^2 +  r^2 d\x^2 \right] +  \frac{ L^2 dr^2}{\f(r) r^2}  \, , 
\stp
where the horizon is at $r=1$ and the boundary is at $r=\infty$. 
$L$ is the AdS radius, $f(r) = 1-1/r^4$, and $T$ is the 
Hawking temperature. 
$r$ is a dimensionless coordinate which measures energy in units of temperature. 
We will also use Eddington-Finkelstein (EF) coordinates to describe the near horizon
dynamics.
In this coordinate system the metric is
\st
    ds^2 = (\pi T)^2 L^2 \left[ - A(r) dv^2 + \frac{2}{\pi T} dr dv  + r^2 d\x^2 \right] \, , 
\stp
where $A=r^2 f(r)$ and $v$ is the EF time 
\st
 v \equiv t  + \frac{1}{\pi T} \int \frac{dr}{fr^2}  \, .
\stp
Ingoing lightlike radial geodesics have $v={\rm const}$, while
outgoing lightlike radial geodesics satisfy $dr/dv =  \pi T A/2$. From now 
on we will set the AdS radius to one, $L=1$.

For simplicity, consider 
fluctuations along an infinitely long straight string ({\it i.e.} an infinitely 
massive quark) stretching from the horizon to the boundary. The stationary 
boundary endpoint is at $x=0$,  
and small fluctuations 
around the straight string solution are 
parameterized by $x(t,r)$, where $x$ denotes displacement
of the string in the $x$ direction. Either $t$ and $r$ or $v$ and $r$ are taken to be the
world sheet parameters, $t=\sigma^0$ and  $r=\sigma^1$. 
The action of these world sheet fluctuations  is derived by linearizing
the Nambu-Goto action:
\st
\label{sng}
S = \frac{\sqrt{\lambda}}{2\pi} \int \dd t\dd r \, g_{xx} \left[ -\frac{1}{2} \sqrt{h} h^{\mu \nu} \partial_{\mu}x \partial_{\nu} x \right] \, ,
\stp
where $\mu,\nu$ run over $t,r$ or $v,r$ depending on the coordinate system.
For example, the world sheet metric in EF coordinates is 
\st
 h_{\mu\nu} \dd \sigma^{\mu} \dd \sigma^{\nu} = (\pi T)^2  
 \left[ - A(r) \dd v^2 + \frac{2}{\pi T} \dd r \dd v \right] \, .   \label{stringmetric}
\stp
We note that the drag coefficient of the heavy quark (see \Eq{langevin}) is related 
to the coupling between the metric and world sheet fluctuations \cite{HKKKY,Gubser,Jorge}
\st
\label{drag_coeff}
 \eta \equiv  \frac{\sqrt{\lambda}}{2\pi} g_{xx}(r_h) = \frac{\sqrt{\lambda}}{2\pi}  (\pi T)^2  \, .
\stp
The analyses in the following sections make this physical interpretation 
of $\eta$ clear. 

The goal of this paper is to show by solving equations of motion 
that 
in equilibrium
the retarded propagator,
\st
    iG_{ra}(t_1r_1| t_2r_2) \equiv \theta(t - t') \llangle \left[ \hat x(t_1, r_1), \hat x(t_2, r_2) \right] \rrangle  \, ,    \label{gret}
\stp
and the symmetrized propagator,
\st
    G_{rr}(t_1r_1| t_2r_2) \equiv \frac12 \llangle \left\{ \hat x(t_1, r_1),  \hat x(t_2, r_2) \right\} \rrangle  \, , 
\stp
are related by the fluctuation dissipation theorem
\st
\label{FDT}
     G_{rr}(\omega, r,\bar r) = -(1 + 2n(\omega)) \,  \Im G_{ra}(\omega, r, \bar r)  \, .
\stp
Here $n(\omega) = 1/(\exp(\omega/T) - 1)$ is the Bose-Einstein distribution function.
This relation is a direct consequence of the fact that the density
matrix is $\exp(-H/T)$
and  signifies
that the fluctuations are in  equilibrium with the black hole at temperature
$T$. 

The advanced propagator is  related to the retarded propagator 
by time reversal,
\st
          iG_{ar}(t_1r_1|t_2r_2) \equiv iG_{ra}(t_2 r_2|t_1r_1) \, ,
\stp
while the spectral density  is  the full commutator
\st
  iG_{ra-ar}(t_1r_1|t_2r_2)  \equiv \llangle \left[\hat x(t_1,r_1),\hat x(t_2,r_2) \right] \rrangle \,.
\stp
Simple manipulations show that
\st
-2\Im G_{ra}(\omega,r,\bar r) = iG_{ra-ar}(\omega,r,\bar r) \,  ,
\stp
and thus  the fluctuation dissipation theorem is a
relation between the commutator   and anti-commutator which 
signifies equilibrium. 
Both the commutator and the anti-commutator will  be determined by solving
equations of motion with appropriate initial conditions. 

\section{Equilibrium string fluctuations in $AdS_5$ from equations of motion}
\label{sec2}

\subsection{Equations of motion and boundary conditions}

Let us analyze the equations of motion in more detail.
Recall that the retarded correlator is a Green function of the equations of motion
\st
\label{eomgreen}
 \frac{\sqrt{\lambda}}{2\pi} \left[ \partial_{\mu} \, g_{xx} \sqrt{h} h^{\mu\nu} \partial_\nu  \right] G_{ra}(t_1r_1|t_2 r_2) = \delta(t_1 - t_2) \delta(r_1 - r_2) \, ,
\stp
and is required to vanish when $t_1r_1$ is outside the future light cone of
$t_2r_2$.  By contrast, the full commutator ({\it i.e.} the spectral density) is not a Green function but satisfies the homogeneous equations of motion
\st
\label{homogeneousra}
 \frac{\sqrt{\lambda}}{2\pi} \left[ \partial_{\mu} \, g_{xx} \sqrt{h} h^{\mu\nu} \partial_\nu  \right] G_{ra-ra}(t_1r_1|t_2 r_2) =  0 \, , 
\stp
where the initial conditions   are determined by the canonical commutation
relations.
Similarly, the symmetrized correlation function 
also satisfies the homogeneous equations of motion
\st
\label{homogeneous}
 \frac{\sqrt{\lambda}}{2\pi} \left[ \partial_{\mu} \, g_{xx} \sqrt{h} h^{\mu\nu} \partial_\nu  \right] G_{rr}(t_1r_1|t_2 r_2) =  0 \, ,
\stp
but the initial conditions  are determined by the density matrix of the
quantum system far in the past. 
The appropriate initial conditions for $G_{rr}$ and $G_{ra-ar}$ are
discussed more fully in \Sect{propagating}. 
Finally all bulk to bulk correlation functions ($G_{ra},G_{ra-ar},G_{rr}$) satisfy Dirichlet, or normalizable, boundary conditions for asymptotically large radius, {\it i.e.}   $G\rightarrow 0$  for $r_1,r_2 \rightarrow \infty$.

Since the supergravity equations of motion are essentially 
coupled oscillators, 
it is useful to recognize that the retarded
propagator for the simple harmonic oscillator is independent of the density matrix. Only symmetrized
correlations  depend on the density matrix and reveal a thermal state.
Since the simple harmonic oscillator clearly illustrates the role of the density
matrix, we show how to compute commutator and anti-commutator oscillator
correlations  using the Keldysh formalism  in \app{shoapp}.

\subsection{Horizon Correlators}
\label{strategy}

The equations of motion propagate initial data in the past  to the future.  This can 
be made manifest for the symmetrized correlator by writing down a formal solution
to Eq.~(\ref{homogeneous}) in terms of retarded correlators.  Specifically, given $G_{rr}$ and its time derivatives
on some time slice $t_1 = t_2 = t_0$, the solution to Eq.~(\ref{homogeneous}) at later times is given by 
\begin{multline}
G_{rr}(1|2) =
 \left[ \frac{\sqrt{\lambda}}{2\pi}\int \dd r_1' \; g_{xx}\sqrt{h}h^{tt}(r_{1}')\, G_{ra}(1|1') \overleftrightarrow{\partial_{t_1'}}\right]  \\
\times \left[ \frac{\sqrt{\lambda}}{2\pi} \int \dd r_2' \; g_{xx}\sqrt{h}h^{tt}(r_{2}')\, G_{ra}(2|2') \overleftrightarrow{\partial_{t_2'}} \right]
G_{rr}(1'|2')  \, , \label{formal}
\end{multline}
where $t_1'$ and $t_2'$ are set equal to $t_0$ after differentiating, and $\ddiff = \overrightarrow{\partial}-\overleftarrow{\partial}$.
This formula expresses the uniqueness of the correlator given its value and time derivatives on a Cauchy surface.%
\footnote
  {
  The formula manifestly satisfies the equations of motion (\ref{homogeneous}) for $t_1,t_2 > t_0$.  To 
  see that it satisfies the  boundary conditions in the limit $t_1 \to t_0$,
  one must know the time derivatives of $G_{ra}(1|1')$  for $t_1\to t_1'$.  This
  derivative
  can  be obtained 
  by using the fact that $G_{ra}(1|1')$ vanishes for $t_1 < t_1'$, and by integrating $t_1$ across $t_1'$ with
  the equations of motion (\ref{eomgreen}) 
  to yield
  the canonical commutation relations
\[
  \lim_{t_1\to t_1'}  \frac{\sqrt{\lambda}}{2\pi} g_{xx} \sqrt{h} h^{tt} \partial_{t_1} G(1|1') = \delta(r_1 - r_1' ) \, .
\]
  An analogous formula holds for the derivative with respect to $t'_1$.
  }
The physical interpretation 
of this solution is easy to understand.  The two retarded Green functions appearing in the integrals 
are convoluted with the separate arguments of the initial data.  These retarded Green functions causally
propagate the initial data forward in time.

\begin{figure}
\begin{center}
\includegraphics[width=0.9\textwidth]{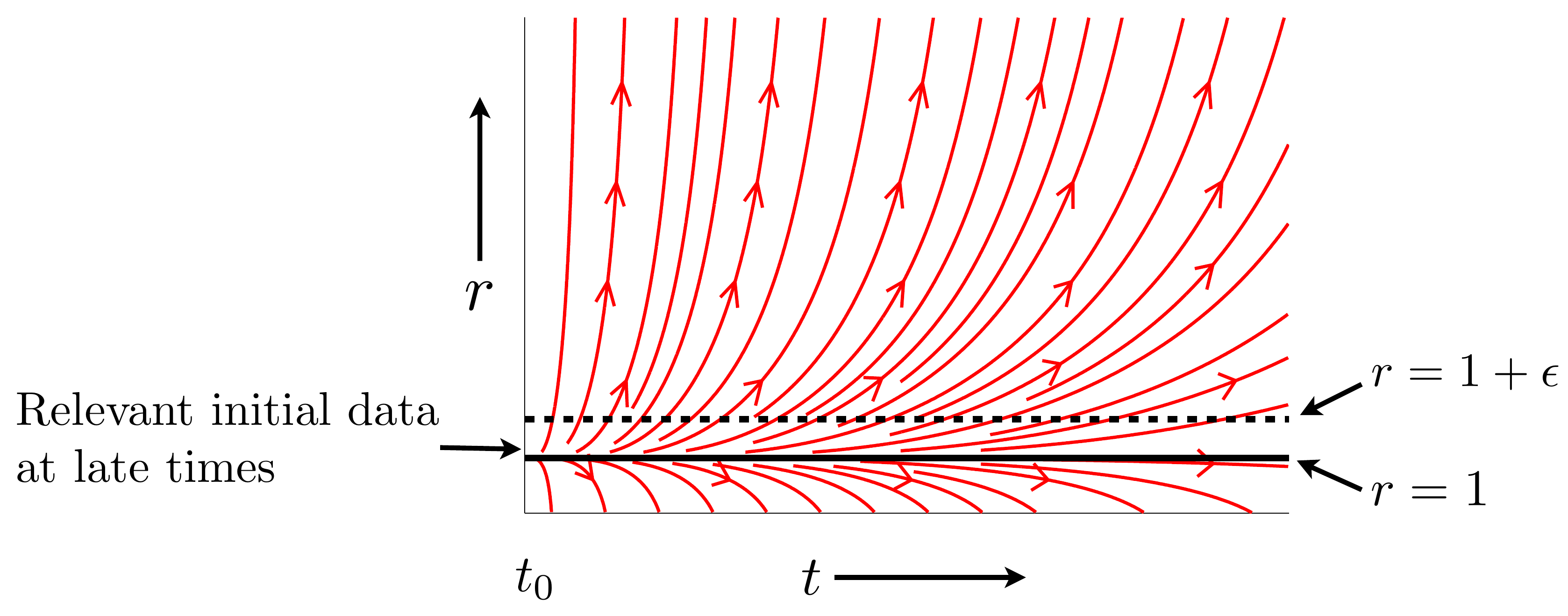}
\end{center}
\caption{ A congruence of outgoing null radial geodesics starting at time $t_0$.  Qualitative 
insight on the propagation of initial data specified on the slice at $t = t_0$ can be understood
from the congruence.  Generic geodesics reach the boundary in a time $\Delta t \sim 1/T$.
Initial data propagated on such trajectories reflects from the boundary and falls into the horizon
with an infall time of order $1/T$.
Geodesics originating exponentially close to the horizon take much longer to escape.
Consequently, at late times the above-horizon geometry is filled with geodesics emanating 
exponentially close to the horizon at $t = t_0$.  Because of this, the only initial data relevant 
at late times consists of the initial data exponentially close to the horizon at $t = t_0$.
\label{geodesics_markup}
}
\end{figure}

To gain  qualitative insight into how initial data is propagated by the retarded Green functions,
Figure~\ref{geodesics_markup} shows a congruence of outgoing radial null geodesics starting at time $t = t_0$.  A generic
geodesic reaches the boundary in a time $\Delta t \sim 1/T$.  The information which is propagated along such
trajectories reflects off the boundary and falls into the black hole with an infall time also of order $1/T$.
Thus, at times $\Delta t \gg 1/T$, the only outgoing geodesics which populate the 
geometry above the stretched horizon
are those which started exponentially close to the horizon at $t = t_0$.
Moreover,  the initial data propagated from this 
exponentially narrow strip to the above-horizon geometry will be dramatically redshifted.  
Because of the redshift, the only finite wavelength 
contributions to the symmetrized correlator will come from the UV part of the 
initial data near the horizon.  This UV part simply comes from 
coincident point singularities in $G_{rr}$, which encode quantum fluctuations in the exponentially narrow strip
near the event horizon, $r=1$. 

This discussion suggests a strategy for computing the symmetrized correlator
at times long after $t_0$.  First, all initial data above the stretched horizon
at $r_h \equiv 1+ \epsilon$ can be neglected as it gets absorbed 
by the black hole at times $t \sim t_0 + 1/T$.  Initial data which lies in the strip between the
event horizon and the stretched horizon can be evolved until it reaches
the stretched horizon, where it  determines  an effective horizon correlator. This
correlator  encodes quantum fluctuations in the past and sources radiation  on the stretched horizon that subsequently propagates up to the
boundary.  This is depicted graphically below in Figure~\ref{geo2_markup}.
Given the discussion of the preceding paragraph, we expect that the form of the
horizon correlator will be independent of details of the initial data specified
in the distant past.

The effective horizon correlation function can be found 
by exploiting the composition law obeyed by retarded propagators.  
Let $g_{ra}(t_1r_2|t_2r_2)$ denote the retarded propagator in the region $1 \le r \le 1+\epsilon$
subject to the reflective Dirichlet condition%
\footnote
  {
  We note that it is not necessary that $g_{ra}$ satisfy reflective Dirichlet condition
  at $r = r_h$.  This \textit{choice} is simply a matter of latter convenience.
  }
at $r = 1+\epsilon$
\st
    g_{ra}(t_1 r_1 = r_h | t_2 r_2) = 0\, ,  \qquad g_{ra}(t_1r_1|t_2 r_2 = r_h) = 0 \, . 
\stp
\app{wronskianapp} analyzes the  Wronskian of the retarded propagators $G_{ra}$ and $g_{ra}$, and 
determines the following composition rule
\st
 G_{ra}(1|1') = \int \dd t_{2} \, G_{ra}(1|2) \left[ \eta \sqrt{h} h^{rr}(r_2)\partial_{r_{2}} \right]_{r_2=r_h}  g_{ra}(2 |1') \, , \label{composition}
\stp
where $t_1r_1$ is outside the stretched horizon, while $t_1'r_1'$ is inside the
stretched horizon.  This identity, schematically depicted in
\Fig{geo2_markup}, is the mathematical statement of how information is propagated up from near the horizon to the stretched
horizon, and then up from the stretched horizon towards the boundary. 
Substituting this composition law into the solution (\ref{formal}) and neglecting contributions 
to the integrals from $r \ge 1 + \epsilon$, we find that above the stretched horizon $G_{rr}$ takes the form
\st
 G_{rr}(t_1r_1|t_2r_2) = \int \dd t_1' \dd t_2' \, \left[-G_{ra}(t_1r_1|t_1'r_h) \right] \; \left[-G_{ra}(t_2r_2|t_2'r_h) \right] \; G_{rr}^h(t_1'|t_2') \, ,
 \label{convolution}
\stp
where the ``horizon correlator'' $G_{rr}^h$ 
is determined only by the dynamics between the horizon and the stretched
horizon:
\st
 G_{rr}^h(t_1|t_2) =  \left[ -\eta \sqrt{h} h^{rr}(r_1)  \partial_{r_1} \right]  \left[ -\eta \sqrt{h} h^{rr}(r_2) \partial_{r_2} \right] g_{rr}(t_1r_1|t_2 r_2) \Big |_{r_1, r_2 = r_h}\,.
 \label{GrrH}
\stp
Here $g_{rr}(t_1r_2|t_2 r_2)$ is the solution to the homogeneous
equations of motion (\ref{homogeneous}) with reflective Dirichlet boundary conditions at the stretched horizon, 
together with the prescribed initial conditions close to the horizon at $t=t_0$.%
\footnote
  {
  In particular, $g_{rr}(1|2)$ is given by the same 
  expression in Eq.~(\ref{formal}), but with the replacements $G_{rr}(1|2) \to g_{rr}(1|2)$,
  $G_{ra}(1|2) \to g_{ra}(1|2)$ and with the limits of integration running from $1 \le r \le 1 + \epsilon$.
  }
We will always denote bulk correlation functions inside the strip
with lower case letters.%
\footnote
  {
  This correlator will always be 
  written with its arguments $g_{rr}(v_1r_1|v_2r_2)$,
  and can not be confused with  the metric coefficient $g_{rr}(r,v)$. 
  }

While the symmetrized correlator measures the degree of occupation of microstates, the spectral density $iG_{ra-ar}(\omega)$
measures the density of available states.  
Since the spectral density is also  a solution to the homogeneous equations of
motion, the discussion of the previous paragraph can be repeated {\it mutatis,
mutandis} yielding
\st
 iG_{ra-ar}(t_1r_1|t_2r_2) = \int \dd t_1' \dd t_2' \, \left[-G_{ra}(t_1r_1|t_1'r_h) \right] \, \left[-G_{ra}(t_2r_2|t_2'r_h) \right] \, iG_{ra-ar}^h(t_1'|t_2') \, ,
 \label{convolutionspec}
\stp
where the horizon spectral density $G_{ra-ar}^h$ is related to the spectral density inside strip $g_{ra-ar}$
 as in \Eq{GrrH}
\begin{equation}
 G_{ra-ar}^h(t_1|t_2) =  \left[ -\eta \sqrt{h} h^{rr}(r_1)  \partial_{r_1} \right]  \left[ -\eta \sqrt{h} h^{rr}(r_2) \partial_{r_2} \right] g_{ra-ar}(t_1r_1|t_2 r_2) \Big |_{r_1, r_2 = r_h}\, .
 \label{GraarH}
\end{equation}
In this equation $g_{ra-ar}$ is subject to the same reflective Dirichlet boundary conditions at $r=1+\epsilon$. 
We will see that the horizon correlators, $G_{ra-ar}^h$ and $G_{rr}^h$, form
the components of a horizon effective action, which  we will  study more
completely in \Sect{HorizonEFT}.

\Eq{convolution} and \Eq{convolutionspec} are the key equations in our study of
thermalization.  Together they  show that emission and absorption of
fluctuations  in the bulk can be encapsulated into a horizon fluctuation and a
horizon resistance.  These equations show that we can focus our attention on
the dynamics inside a small strip between the horizon and stretched horizon.
Thermalization inside the strip will be simple to understand analytically.

\begin{figure}
\begin{center}
\includegraphics[width=0.75\textwidth]{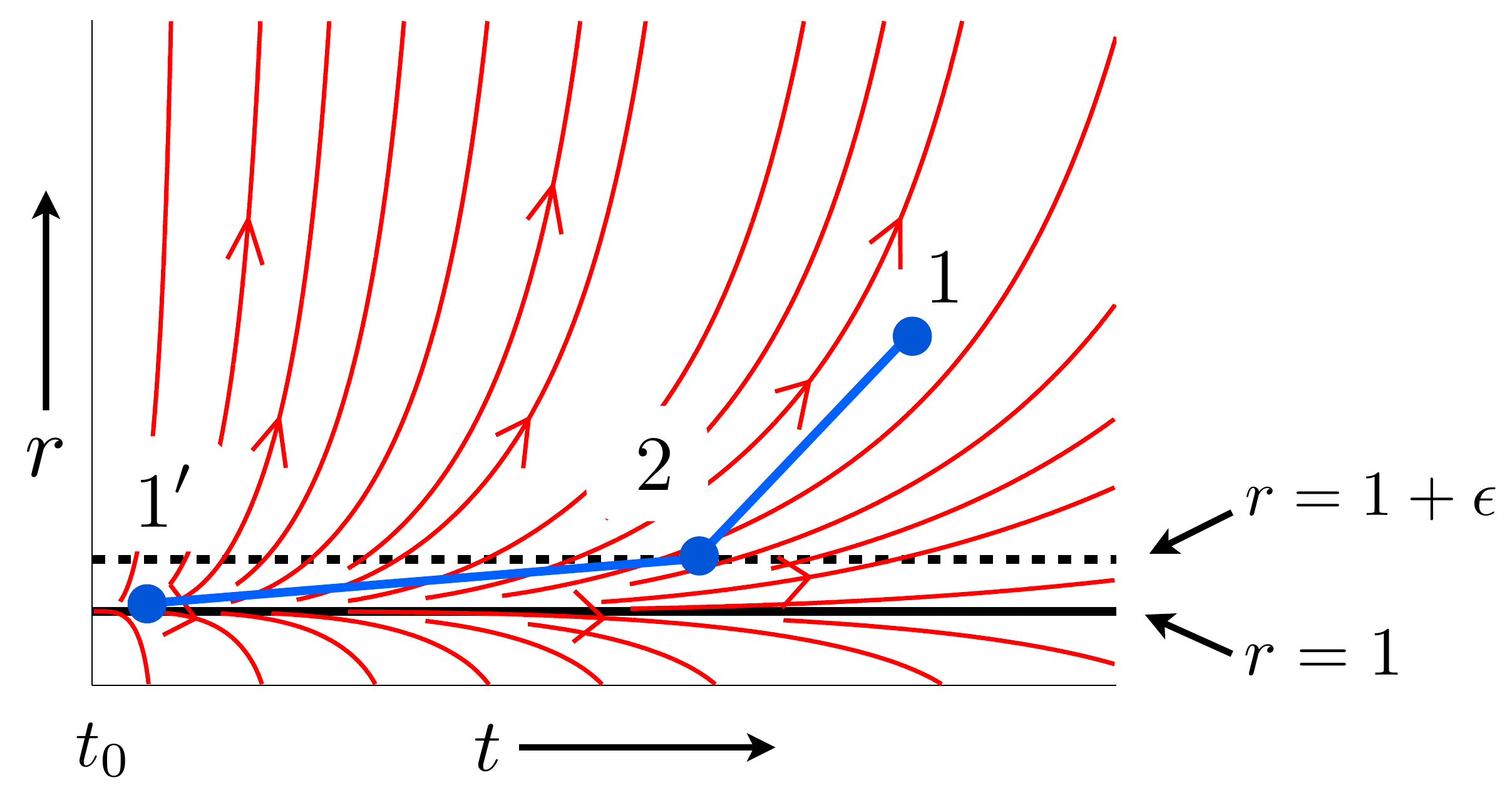}
\end{center}
\caption{ The composition law for retarded Green functions.  The stretched horizon separates the two steps of the evolution.  From the perspective of an observer in the exterior,
the strip between the horizon at $r = 1$ and the stretched horizon at $r = 1 + \epsilon$ produces a simple horizon effective correlator, which acts like a 
source of radiation from the stretched horizon.
\label{geo2_markup}
}
\end{figure}

\subsection{Propagating correlations to the stretched horizon }
\label{propagating}

\subsubsection{ The horizon symmetrized correlation }
\label{sec:horizonsymcorr}

As is evident from Eqs.~(\ref{convolution}) and (\ref{GrrH}), to determine the symmetrized correlator
we must first determine $g_{rr}(t_1r_1|t_2 r_2)$ in the strip between the horizon and the stretched horizon.
For describing the evolution inside the strip it is useful to use Eddington-Finkelstein coordinates.
In these coordinates the equations of motion for $r < r_h$ are
\st
\label{eq1grr}
 \frac{\partial}{\partial r_1}  \left[ 2 h^{v r}\frac{\partial}{ \partial v_1} +    h^{rr}(r_1) \frac{\partial}{\partial r_1} \right] 
g_{rr} (v_1r_1|v_2r_2) = 0   \, ,
\stp
and
\st
\label{eq2grr}
 \frac{\partial}{\partial r_2}  \left[ 2 h^{v r} \frac{\partial}{ \partial v_2} +    h^{rr}(r_2) \frac{\partial}{\partial r_2} \right] 
g_{rr}(v_1r_1|v_2r_2) = 0  \, , 
\stp
where the metric coefficients are given in \Eq{stringmetric}.
Since we are interested in the ultra-violet  irregular solution to \Eq{eq1grr} and \Eq{eq2grr}, 
we have neglected  the radial
derivatives of $g_{xx}$, which are small compared to the radial
derivatives of $g_{rr}(v_1 r_1|v_2r_2)$.%
\footnote{A posteriori one can verify that the neglected derivatives, {\it e.g.}
$(\partial_r g_{xx})\,(\partial_v g_{rr}(v_1 r_1|v_2r_2))$,   are
small compared to the terms which are kept, {\it e.g.} $g_{xx} \, \partial_r h^{vr} \partial_v g_{rr}(1|2)$. }
With this short-distance approximation, the linear operator in \Eq{eq1grr}
becomes a product of two operators, 
$\partial_r$  and $\partial_v + \frac{\pi T}{2} A\,  \partial_r$.
Recall that in Eddington-Finkelstein coordinates, outgoing null radial geodesics satisfy $\frac{dr}{dv} = \frac{\pi T}{2} A$, 
whereas infalling null geodesics satisfy $v = $ const.
Thus, the derivatives act along these geodesics,
implying that functions of  $e^{-2 \pi T v}(r - 1)$ and 
$v$ satisfy the equations of motion in the near horizon limit.  The general solution is therefore
\begin{align}
g_{rr}(v_1r_1|v_2r_2) =  f_1(e^{-2 \pi T v_1}(r_1 - 1),e^{-2 \pi T v_2}(r_2 - 1)) 
+f_2(e^{-2 \pi T v_1}(r_1 - 1),v_2) \\ \nonumber
+\ f_3(v_1,e^{-2 \pi T v_2}(r_2 - 1)) 
+f_4(v_1,v_2),
\end{align}
where the $f_n(x_1,x_2)$ are arbitrary functions.  Requiring that the boundary conditions
\[
g_{rr}(v_1,r_1 = r_h| v_2r_2) = 0\, , \quad \mbox{and} \quad 
g_{rr}(v_1r_1 | v_2,r_2 = r_h) = 0  \, , 
\]
are satisfied at all times, we find  that
\begin{align}
\label{feqn}
g_{rr} (v_1r_1|v_2r_2) =&  +f( e^{-2\pi T v_1}(r_1 - 1), 
e^{-2\pi T v_2} (r_2 -1 ) ) \nonumber \\
& - f(e^{-2\pi T v_1} (r_1 - 1) , e^{-2\pi T v_2} (r_h - 1)) \nonumber \\
& - f(e^{-2\pi T v_1} (r_h - 1) , e^{-2\pi T v_2} (r_2 - 1)) \nonumber \\
& + f(e^{-2\pi T v_1} (r_h - 1) , e^{-2\pi T v_2} (r_h - 1))  \, ,
\end{align}
where $f(x_1, x_2)$ is determined by initial conditions. 

The above solution is a linear combination of modes which are 
outgoing from the horizon and modes which are infalling towards the horizon.
The infalling modes are in fact a consequence of the outgoing modes, as the 
reflective Dirichlet boundary conditions at the stretched horizon turn
outgoing modes into infalling modes, which are subsequently absorbed by the black hole.
Inspection of \Eq{feqn} shows that 
if $f(x_1,x_2)$ is analytic, this reflection and absorption would always
lead to $g_{rr}(v_1r_1|v_2r_2) \to 0$ at late times.
Furthermore, all outgoing modes originate exponentially close
to the horizon.  This follows from the fact that as $e^{-2 \pi T v}(r - 1)$ is constant
on outgoing null geodesics, in the limit $v \to - \infty$ we must have $r \to 1$.
Therefore, for evolution near $v = 0$, relevant initial data specified in the distant past 
will come from an exponentially narrow strip which is exponentially close to the horizon
at $r = 1$.
%
%
In this exponentially narrow strip $g_{rr}(v_1r_1|v_2r_2)$
is not analytic and contains a logarithmic coincident point singularity 
\begin{equation}
\label{coincidentpts}
g_{rr}(v_1r_1|v_2r_2) =   -\frac{1}{4\pi \eta} \log
\left|  \mu  (v_1 - v_2)(r_1 - r_2)\right| + {\rm exponentially \ small \ terms}, 
\end{equation}
where $\mu$ is a constant.\footnote{The coincident point limit can be found be computing the anti-commutator  for a
free massless scalar field theory  in  1+1d flat space 
\[
 \frac{1}{2} \llangle \left\{\phi(X) , \phi(0) \right\} \rrangle = -\frac{1}{4\pi} \log|\mu \,  \eta_{\mu\nu} X^{\mu} X^{\nu}  | \, .
\]
\Eq{coincidentpts} is found by the principle of equivalence and 
by comparing the normalization of the string action (\ref{sng})  to the canonical normalization.
  } 

By matching Eq.~(\ref{feqn}) on to Eq.~(\ref{coincidentpts}) we conclude that we must have
\begin{equation}
\label{fasym}
f(x_1, x_2)  =  -\frac{1}{4\pi \eta} \log|x_1 - x_2| + {\rm exponentially \ small \ terms}.
\end{equation}
Substituting Eq.~(\ref{fasym}) into Eq.~(\ref{feqn}), we find that long after initial conditions were specified 
and up to exponentially small corrections, we have
\begin{align}
\label{grrstrip}
g_{rr} (v_1r_1|v_2r_2) =&  -\frac{1}{4\pi \eta} \log| e^{-2\pi T \Delta v} (r_1 {-} 1)  -  (r_2 {-} 1) |  + \frac{1}{4\pi \eta}  \log |e^{-2\pi T \Delta v} (r_1 {-} 1)  -  (r_h {-} 1)| \nonumber \\
& + \frac{1}{4\pi \eta}  \log | e^{-2\pi T\Delta v} (r_h {-} 1)  -  (r_2 {-} 1)|  - \frac{1}{4\pi \eta}  \log |e^{-2\pi T \Delta v} (r_h{-} 1) - (r_h {-} 1)|  \, .
\end{align}
In writing \Eq{grrstrip} we have pulled  out a common factor of $e^{-2\pi T v_2}$ which cancels between the different terms due to the reflective boundary conditions.


The fact that $g_{rr}(v_1r_1|v_2r_2)$ is time translationally invariant at late times 
is due to the  competition between  emission 
and absorption.
Quantum fluctuations give rise to the logarithmic coincident point singularity 
in $g_{rr}(v_1r_1|v_2r_2)$, and this singularity acts as a constant source 
of radiation.  Thus, radiation is continually produced, redshifted, reflected off the stretched horizon, and absorbed by the black hole.  
This dynamical process 
reaches a steady state  which  determines $g_{rr}(v_1r_1|v_2r_2)$ at late times.  
As we will discuss in Section~\ref{sec:bulkfdteom},
this steady state is in fact a thermal state.

With $g_{rr}(v_1r_1|v_2r_2)$ known, we may compute the horizon correlator $G_{rr}^h(v_1,v_2)$ via Eq.~(\ref{GrrH}).  The result
reads
\begin{align}
G_{rr}^h (v_1,v_2) =&   \label{symg}
   -\frac{\eta}{\pi} \partial_{v_1} \partial_{v_2}  \log | 1 - e^{-2\pi T (v_1 - v_2) } | \, .
\end{align}




%

%


%

\subsubsection{The horizon spectral density}
\label{sec:horizonspectraldensity}

As in the case of $g_{rr}(t_1r_1|t_2r_2)$, 
it is convenient to use EF coordinates $(v,r)$ to determine $g_{ra-ar}(t_1r_1|t_2r_2)$ .
The spectral density $g_{ra-ar}(v_1r_1|v_2r_2)$ obeys the same equation of motion as $g_{rr}(v_1r_1|v_2r_2)$ and has the same reflective boundary conditions
at the stretched horizon. Thus, the general solution for 
$g_{ra-ar}(v_1r_1|v_2r_2)$ is similar to the corresponding solution for $g_{rr}(v_1r_1|v_2r_2)$ in Eq.~(\ref{feqn}).  
Explicitly, we have
\begin{align}
 g_{ra-ar}(v_1r_1|v_2,r_2) =&  +\otherf( e^{-2\pi T v_1}(r_1 - 1), 
e^{-2\pi T v_2} (r_2 -1 ) ) \nonumber \\
& - \otherf(e^{-2\pi T v_1} (r_1 - 1) , e^{-2\pi T v_2} (r_h - 1)) \nonumber \\
& - \otherf(e^{-2\pi T v_1} (r_h - 1) , e^{-2\pi T v_2} (r_2 - 1)) \nonumber \\
& +\otherf(e^{-2\pi T v_1} (r_h - 1) , e^{-2\pi T v_2} (r_h - 1)) \, , \label{spectral}
\end{align}
where $\otherf$ is determined by initial conditions.

In contrast to the symmetrized correlator, where the initial conditions  
are determined by the density matrix in the past,
the initial conditions for the spectral density are state independent  
and are determined by the canonical commutation relations.  
Thus, it is not necessary to evolve for a long time 
before reaching a steady state solution.

For $v_1r_1$ close to $v_2r_2$ flat-space physics determines 
$g_{ra-ar}$  \footnote{
The coincident point limit can be found by computing the commutator 
for a free massless scalar field theory  in  1+1d flat space 
\[
 -i\llangle [\phi(X) , \phi(0)] \rrangle = -\frac{1}{2} \theta(-\eta_{\mu\nu}
X^{\mu} X^{\nu} ) \, \sign(t) = \frac{-1}{4} ({\rm sign}(t + z) -  {\rm
sign}(-t+z)  ) \, .
\]
The sign function is the only functional form which allows left-moving and right-moving modes to cancel at spacelike separations.
\Eq{shortdistspec} is found by the principle of equivalence and   
by comparing the normalization of the string action (\ref{sng})  to the canonical normalization. 
}
\begin{equation}
\label{shortdistspec}
  g_{ra-ar}(v_1 r_1| v_2 r_2)   
\rightarrow
    -\frac{1}{4 \eta} \left (\sign(v_1 - v_2) - \sign(r_1 - r_2) \right ) \, .
\end{equation}
Comparing Eq.~(\ref{shortdistspec}) to Eq.~(\ref{spectral}), we conclude that for any $v_1$ and $v_2$ we must 
have
\st
\label{signeqn}
  f_{ra-ar}(x_1,x_2) = \frac{1}{4\eta} {\rm sign} (x_1 - x_2) \, .
\stp 
The infalling and outgoing modes cancel at spacelike separation due to the 
sign function.
One can verify (with careful algebra) that the $g_{ra-ar}$ satisfies
the canonical commutation relation
\st
    \eta \sqrt{h} h^{tt}(r_1) \lim_{\rm t_2 \to t_1} \partial_{t_1}g_{ra-ar}(t_1r_1|t_2r_2) =  \delta(r_1 - r_2)  \, .
\stp

Substituting $g_{ra-ar}$ into Eq.~(\ref{GraarH}),
we determine the horizon spectral density
\st
  G_{ra-ar}^h(v_1,v_2)   \label{specg}
   =  2 \eta \, \delta'(v_1-v_2)\,.
\stp

\subsection{The Bulk Fluctuation Dissipation Theorem}
\label{sec:bulkfdteom}

The interpretation of the above results becomes clear in Fourier space.  The Fourier transform of the horizon symmetrized correlator is
\st
 G_{rr}^h(\omega) =  \frac{\eta}{\pi} \int_{-\infty}^\infty \dd v \, e^{i\omega v}\,  \partial_v^2 \log| 1 - e^{-2\pi Tv}|  = \eta\omega\, (1+2n(\omega)) \, ,  \label{bdygrr}
\stp
where $n(\omega) = 1/(\exp(\omega/T) - 1)$ is the Bose-Einstein distribution.  The Fourier transform of the horizon spectral density is
\st
 iG_{ra-ar}^h(\omega) = 2\eta\omega\, .
\stp
Thus, after the decays of transients in the initial data, the horizon correlation functions obey the fluctuation dissipation relation
\st
 G_{rr}^h(\omega) = iG_{ar-ar}^h(\omega)\left(\frac{1}{2} +n(\omega)\right) \, .  \label{horizonfluct}
\stp

\begin{figure}
\begin{center}
\includegraphics[width=0.45\textwidth]{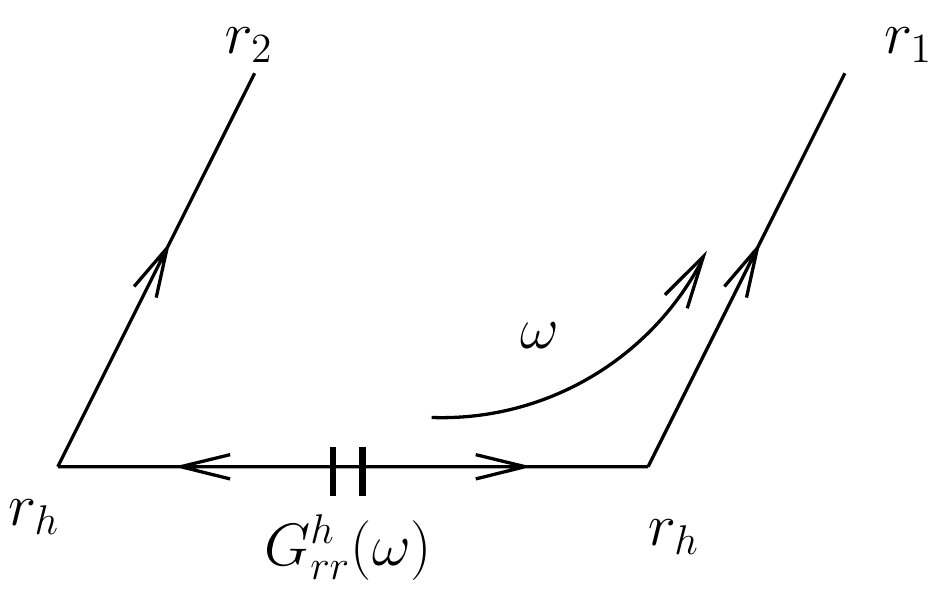}
\end{center}
\caption{
\label{horizon_graph}
Feynman graph used for computing the symmetrized correlation 
function $G_{rr}(\omega, r_1, r_2)$, see \protect \Eq{gsym}.  }
\end{figure}

Now we can see  how the bulk will thermalize from these horizon correlations \cite{Son:2009vu}.
In Fourier space the expression for $G_{rr}$ in terms of $G_{rr}^h$ in Eq.~(\ref{convolution}) becomes
\begin{align}
\label{gsym}
  G_{rr}(\omega, r_1, r_2) =& G_{ra}(\omega,r_1,r_h) G_{ra}(-\omega,r_2,r_h) \left[ G_{rr}^h(\omega) \right] \, ,
\end{align}
and the spectral density obeys a similar equation
\st
 iG_{ra-ar}(\omega, r_1, r_2) = G_{ra}(\omega,r_1,r_h) G_{ra}(-\omega,r_2,r_h)
\left[ iG_{ra-ar}^h(\omega)  \right] \, . \label{spectraldensity}
\stp
Thus the horizon fluctuation dissipation relation 
trivially implies  the relation in bulk
\st
 G_{rr}(\omega, r_1, r_2) = iG_{ra-ar}(\omega, r_1, r_2) \left(\frac12 + n(\omega) \right) \, .
\stp
In \Sect{HorizonEFT} we will show how the convolution in \Eq{gsym}  
is the result of coupling the  horizon effective action to the 
bulk.  Anticipating these results, 
\Fig{horizon_graph} shows the Feynman graph corresponding to \Eq{gsym}.


Physically, the bulk fluctuation dissipation theorem follows from its horizon
counterpart because any fluctuation in the bulk must have crossed the stretched horizon
at some point in the past.  Previously the form of the horizon correlators was
derived either by assuming equilibrium, or by  using a complex set of analytic
continuations.  We see that it is the endpoint of a simple competitive dynamics
inside the strip.

\section{The Horizon Effective Action }
\label{HorizonEFT}

It is enlightening and useful to package the above manipulations into
a path integral formalism.
The claim is that these steps correspond to integrating out the fields inside the stretched horizon.  The horizon correlation functions defined above
are the components of the resulting effective action.

\subsection{The Keldysh Formalism}
\label{Keldysh_intro}

\begin{figure}
\includegraphics[width=.6\textwidth]{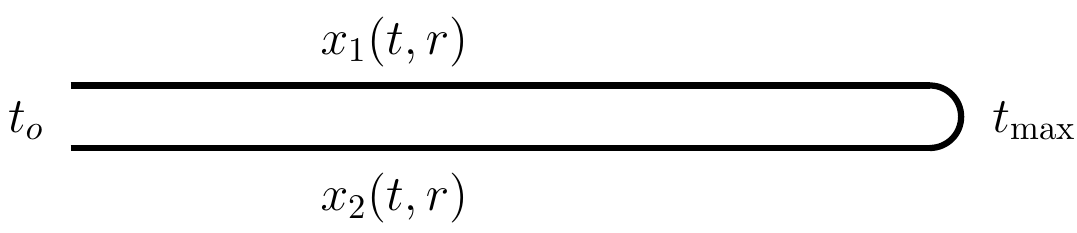}
\caption{\label{contour} 
The Keldysh contour from an initial time $t_{o}$ where the density 
matrix is specified to a final time, $t_{\rm max}$. 
$x_{a}(t,r) = x_1(t,r) - x_2(t,r)$ must be zero at $t_{\rm max}$.
}
\end{figure}

In any quantum statistical system, correlations are computed 
by tracing the density matrix $\rho$  over the
Heisenberg operators. This is the entire purpose of the Keldysh contour  \cite{Chou:1984es}.
The generating function of string correlation functions in bulk is 
\st
\label{exact_generator}
Z[\F_1,\F_2] = \Tr\left[\rho \, \int_{x_1(t_o,r) \atop x_2(t_o,r)} 
\negthinspace 
[\Db x_1] [\Db x_2] \; 
e^{iS_1 - iS_2} \; 
e^{i \int \dd t\dd r   \F_1(t,r)\, x_1(t,r)} \, e^{-i\int \dd t' \dd r'  \, 
\F_2(t',r') \, x_2(t',r') } \right] \nc
\stp
where  $[\Db x_1]$ indicates a bulk path integral, {\it i.e.}
\st 
[\Db x_1] = \prod_{t,r} \dd x_1(t,r)  \, .
\stp
The path integral is defined along the Schwinger-Keldysh contour shown in \Fig{contour}, where the 
``1" type path integral is the amplitude of the process, while the ``2" type
path integral is the conjugate amplitude of the process. The trace is
over the initial density matrix $\rho$ which determines the initial conditions
$x_1(t_o,r)$ and $x_2(t_o,r)$.
$\F_1$ and $\F_2$ are sources, which in this case are simply the external forces applied to the string.
Variation of generating function with respect to $\F_1$ and $\F_2$ 
yields time ordered,  anti-time ordered, and Wightman correlation functions  \cite{Son:2009vu}.

Instead of using the 12 variables, we will rely on a rewritten version of the Keldysh formalism known as the 
$ra$ formalism which is dramatically simpler.
We define  retarded ($r$) and advanced ($a$) fields and sources
\st
  x_{r} = \frac{x_1 + x_2}{2} \nc  \qquad x_a = x_1- x_2 \nc
  \qquad \qquad \F_{r} = \frac{\F_1 + \F_2}{2} \nc \qquad \F_a = \F_1 - \F_2 \nc
\stp
Since $x_a$ encodes the 
 the differences between the amplitude and conjugate amplitudes,
$x_a$ is a small parameter in the classical limit \cite{Mueller:2002gd}.
In terms of $r$ and $a$ fields the action becomes 
\st
\label{actions1s2}
S_1 - S_2 =    \frac{\sqrt{\lambda}}{2\pi} \int \dd t\dd r \, g_{xx} \left[ -\sqrt{h} h^{\mu \nu} \partial_{\mu}x_r \partial_{\nu} x_a \right] \,.
\stp
The two point
functions in the ``ra" formalism are familiar and explain the notation of
\Sect{prelim}
\begin{align}
 iG_{ra}(t_1r_1 | t_2 r_2) =& \llangle x_{r}(t_1,r_1)  x_{a}(t_2, r_2 ) \rrangle = \theta(t_1-t_2) \llangle \left[ \hat x(t_1,r_1),  \hat x(t_2, r_2)  \right] \rrangle \, , \\
 G_{rr}(t_1r_1 | t_2r_2) =& \llangle x_{r}(t_1,r_1)  x_{r}(t_2, r_2 ) \rrangle =  \frac{1}{2} \llangle \left\{ \hat x(t_1,r_1),  \hat x(t_2, r_2)  \right\} \rrangle \, .
\end{align}

The causal structure of quantum field theory is rendered transparent in  $ra$ formalism.
 Since at $t_{\rm max}$ along the contour, $x_1(t_{\rm max})=x_2(t_{\rm max})$,
the path integral must be solved with the boundary condition, $x_{a}(t)
\rightarrow 0$ for $t \rightarrow \infty$. 
Thus, whenever an $``a"$ type field is evaluated at a later time than all other
field insertions the correlator vanishes. For example, $G_{aa}(t,t') =0$ since
an $a$ field is always evaluated last.
 Similarly the retarded correlator, $G_{ra}(t,t')$,
vanishes whenever the $a$ field is evaluated at a  later time than the 
$r$ field.
$G_{ra}$ determines (minus) the  retarded linear response to 
a classical force. 
We will exhibit the retarded correlator with an arrow to
indicate the direction of time recorded by this propagator
\[
\begin{minipage}[c]{1.8in}
\includegraphics[width=1.8in]{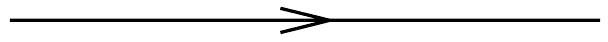} 
\end{minipage}
\quad
\begin{minipage}[c]{1.2in}
$ = iG_{ra}(\omega, r, \bar r) $ \, .
\end{minipage}
\]
While  the retarded correlators reflect the response to a classical force,
symmetrized correlators encode the fluctuations in the system. 
Since the symmetrized correlation function does not
represent causal response to a classical force, but rather a time dependent
correlation which arose from a specified initial condition,    we will notate
this correlation (as in \Ref{CaronHuot:2008uh}) with  
\st
\begin{minipage}[c]{1.8in}
\includegraphics[width=1.8in]{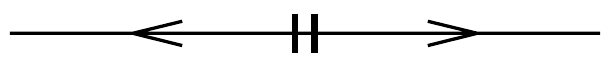} 
\end{minipage}
\quad
\begin{minipage}[c]{1.2in}
$ = G_{rr}(\omega, r, \bar r) $ \, .
\end{minipage}
\stp
Higher point correlation functions involving $r$ and $a$ indices
have a similarly simple interpretation \cite{Chou:1984es}.

\subsection{ The effective action from the path integral}
\label{horizon_effective}

To obtain the horizon effective action we integrate out all field fluctuations inside the stretched horizon
at $r_h=1+\epsilon$. 
The path integral in \Eq{exact_generator}  becomes
\st
\label{zapprox}
Z[\F_1,\F_2] = 
   \int_{r > r_h} \left[\Db x_1 \Db x_2 \right] \,\Dt x^h  \,
e^{iS_1 - iS_2 + iS_{\rm eff}^h[x^h] } \; 
e^{i \int  \F_1 \, x_1} \, e^{-i\int  \, \F_2 \, x_2 }  \nc
\stp
where the horizon effective action is the path integral 
\st
\label{heff}
 e^{ iS_{\rm eff}[x^h_s] } \equiv \Tr\left[\rho \int_{r < r_h \atop {x_s(t_0,r)} } \Db x_s \,  \delta \left[x^h_s(t)  - x_s(r_h, t) \right]  e^{i S_1 - i S_2 } \right] \nc
\stp
with fixed boundary values on the stretched horizon $x^h_s(t)$. 
Here the ``$s$" sub-label denotes the Schwinger-Keldysh index.  The horizon
effective action  must be traced over the initial data extremely close to the
event horizon.  The fixed boundary conditions 
couple the field fluctuations with $r> 1 +\epsilon$ to field fluctuations
with $r< 1 + \epsilon$.
The trace determines how 
vacuum fluctuations in the past  influence  the future dynamics.
It should be understood that \Eq{zapprox} (where the initial density matrix is retained only inside the stretched horizon) is an approximation
that is valid  after most of the initial data has fallen into the hole as discussed in \Sect{sec2}.

To quadratic order the effective action  can be expanded as: 
\st
\label{heff2}
  iS_{\rm \eff}[x^h] = -\int \dd t\, \dd \bar{t} \;  x_{a}^h(t) \left[iG_{ra}(t,\bar t) \right] x_{r}^h(\bar t)  - \frac{1}{2} \int \dd t \, \dd \bar{t} \;  x_{a}^h(t) \left[ G_{rr}^h(t,\bar t)  \right ]  x_{a}^h(\bar t) \, ,
\stp
where we can 
determine the horizon correlators via differentiation, e.g.
\begin{align}
\label{grrrule}
 G_{rr}^h(t,\bar{t}) 
 =& \frac{1}{i^2} \left. \frac{i\,\delta S_{\rm eff}}{ \delta x_a^h(t)  \, \delta x_a^h(\bar t )} \right|_{x_r^h,x_{a}^h = 0} \, , \\
 iG_{ra}^h(t,\bar{t}) 
=& \frac{1}{i^2} \left. \frac{i\delta S_{\rm eff}}{ \delta x_a^h(t) \,  \delta x_r^h(\bar t) } \right|_{x_r^h,x_{a}^h = 0} \, .
\end{align}

A short calculation\footnote{
This is because the only dependence on $x^{h}_s$ comes through the 
boundary terms of the kinetic term, e.g.  
\[
  - \int \dd t\dd r \, \eta \sqrt{h} h^{rr} \partial_r x_r \partial_r x_a = \left. - \eta \sqrt{h} h^{rr} x_{a} \partial_r x_r \right|_{r=r_h} + \int \dd t\dd r \, x_{a} \partial_r\left( \eta \sqrt{h} h^{rr} \partial_r x_r \right) \, .
\]  
}
shows that  every derivative $-i\delta/\delta x_a$  or $-i\delta/\delta x_r$ brings down a factor of $- \eta \sqrt{h} h^{rr} \partial_r x_r$ or $- \eta \sqrt{h}h^{rr} \partial_r x_a$ respectively. Thus 
\begin{align}
\label{horizon_correl}
G_{rr}^h(t,\bar{t}) 
=&  \lim_{r,\bar{r} \rightarrow r_h} \left[-\eta \sqrt{h} h^{rr}(r) \partial_{r} \right] \left[ -\eta \sqrt{h} h^{rr}(\bar r)  \partial_{\bar r} \right] g_{rr}(t_1r_1|t_2r_2) \, ,  \\
G_{ra}^h(t,\bar{t}) 
=&  \lim_{r,\bar{r} \rightarrow r_h} \left[-\eta \sqrt{h} h^{rr}(r) \partial_r \right] \left[-\eta \sqrt{h} h^{rr}(\bar r) \partial_{\bar{r}} \right] g_{ra}(t_1r_1|t_2r_2) \, ,
\end{align}
where the lower case correlation functions are defined from
the path integral for $r < r_h$ with vanishing boundary conditions at $r=r_h$
\begin{align}
\label{path_eff}
 \llangle \ldots \rrangle_{h}
  &=   \frac{1}{Z} \Tr\left[ \rho \int_{r< r_h \atop x_s(t_0, r)} \Db x_s \;  \delta\left[ x_s(v,r_h) \right] e^{iS_1 - i S_2 }\ldots \right] \, .
\end{align}
These relations are familiar from the context of the AdS/CFT correspondence
where one takes radial derivatives of the fields as $r\rightarrow \infty$ 
\cite{Maldacena:1997re,Gubser:1998bc,Witten:1998qj}.
The functions $g_{rr}(t_1r_1|t_2r_2)$ and $g_{ra}(t_1r_1|t_2r_2)$ (or equivalently $g_{ra-ar} = g_{ra} - g_{ar}$) are the same as in section \ref{sec2} and their relation to $G_{rr}^h$ and $G_{ra-ar}^h$ is identical to that derived in \Eq{GrrH}.  Further the reflective
boundary conditions imposed on $g_{rr}$ and $g_{ra}$ appear naturally in the path integral formalism.
We conclude that the horizon correlators $G_{rr}^h$ and $G_{ra-ar}^h$ defined and computed in the previous section are precisely the components of a well-defined effective action. 
Given a procedure for regularizing ultraviolet divergences in gravity, this
effective action could be computed to any desired order in perturbation theory
following the methods of \Ref{symanzik}.

\subsection{Summary of the horizon effective action}
\label{sec:sonteaney}

The effective action in equilibrium was obtained previously by Son and Teaney \cite{Son:2009vu} by analytically continuing modes across the horizon.
Its form is simple,
\st
\label{horizon_act}
iS_{\rm eff}^{h} = -\int \frac{\dd \omega}{2\pi} \, 
x^{h}_a(-\omega) \left[ i G_{ra}^h(\omega)  \right] x^h_r(\omega)  - \frac{1}{2} \int \frac{\dd \omega}{2\pi} x_{a}^{h}(-\omega) \left[ G_{rr}^h(\omega)\right] x_a^{h}(\omega) \nc
\stp
where the retarded horizon correlation function is simply
\st
  G_{ra}^{h} = -i\omega \eta  \, ,
\stp
and the symmetrized part of the  horizon action  obeys
a fluctuation dissipation relation
\st
  G_{rr}^h = - (1 + 2n(\omega)) \, \Im G_{ra}^h(\omega) \, . 
\stp
Fourier transforming back to time, the horizon action reads
\st
 iS_{\rm eff} = -i\int \dd t\,  x_a^h(t)  \,  \eta \partial_t x_r^h(t)  -   \frac{1}{2} \int \dd t \dd t' \,  x_{a}^h(t) \left[ - \frac{\eta}{\pi} \,  
\partial_t \partial_{t'} \log|1 - e^{-2\pi T (t - t')} |
\right] 
x_{a}^h(t') \, .
\stp
This action compactly summarizes all the correlations that appear through
quadratic order in $x_{a}$ and agrees with the results of \Sect{sec:bulkfdteom}.

The horizon effective action is useful. For instance, since the bulk action in \Eq{actions1s2} has no $rr$ and no $aa$ components,  the  first Feynman graph which contributes to
the correlation $\llangle x_r(v_1,r_1) \, x_r(v_2,r_2) \rrangle_h$ is shown in
\Fig{horizon_graph} and is written in full in \Eq{gsym}, with the lines explained in \Sect{Keldysh_intro}. 
Thus, perturbation theory with the bulk and horizon
actions transparently produces the convolution formulas given in \Eq{convolution} and \Eq{convolutionspec}.

The retarded horizon propagator in the action, $G_{ra}^h$, reflects the resistance on the horizon, $-\eta \dot x$, and is valid at all frequencies \cite{Son:2009vu}. 
Indeed, the classical dissipation encoded by the $ra$ part of 
the action can be derived simply from the classical membrane paradigm.
Variation of the effective action gives  the horizon force
\st
 \F^h_r = \frac{\delta S_\eff}{\delta x_a} = -\eta \partial_t x_r \, , 
\stp
where the force is $\F^h_r=-\eta \sqrt{h}h^{rr}\partial_r x_r$ as required by the membrane paradigm \cite{Parikh:1997ma}.
In the zero frequency limit the force $\F_r$
is independent of radius \cite{Iqbal:2008by,HKKKY},  implying 
that horizon drag coefficient $\eta$ is the same as boundary drag coefficient in the Langevin equation \cite{HKKKY}.



\section{Non-equilibrium correlators}

\subsection{Setup}

In this section we wish to show how to generalize the non-equilibrium 
horizon effective action.  Interesting non-equilibrium geometries to consider can be found in \cite{Chesler:2008hg, Chesler:2009cy, Chesler:2010bi}.
However, for definiteness we will consider 
the non-equilibrium geometry discussed in Ref. \cite{Chesler:2008hg}.
In this work, an excited state in the boundary quantum field theory was created by 
briefly turning on a time-dependent $4d$ gravitational field, which was taken to be translationally invariant.  The gravitational 
field did work on the quantum system, producing an excited state which subsequently thermalized.  In the dual
$5d$ gravitational system, turning on a $4d$ gravitational field corresponds to deforming the $4d$ boundary 
of the $5d$ geometry.  Before the deformation was turned on, the $5d$ geometry was taken to be $AdS_5$, which 
is dual to the vacuum state.  The deformation in the boundary geometry produced gravitational radiation which fell into the bulk.
This infalling radiation resulted in the process of gravitational collapse, changing the initial $AdS_5$ geometry to one which had a 
black hole, and the relaxation of the black hole to equilibrium encoded the thermalization 
of the expectation value of the stress tensor in the dual quantum theory. 

\begin{figure}
\begin{center}
\includegraphics[width=0.5\textwidth]{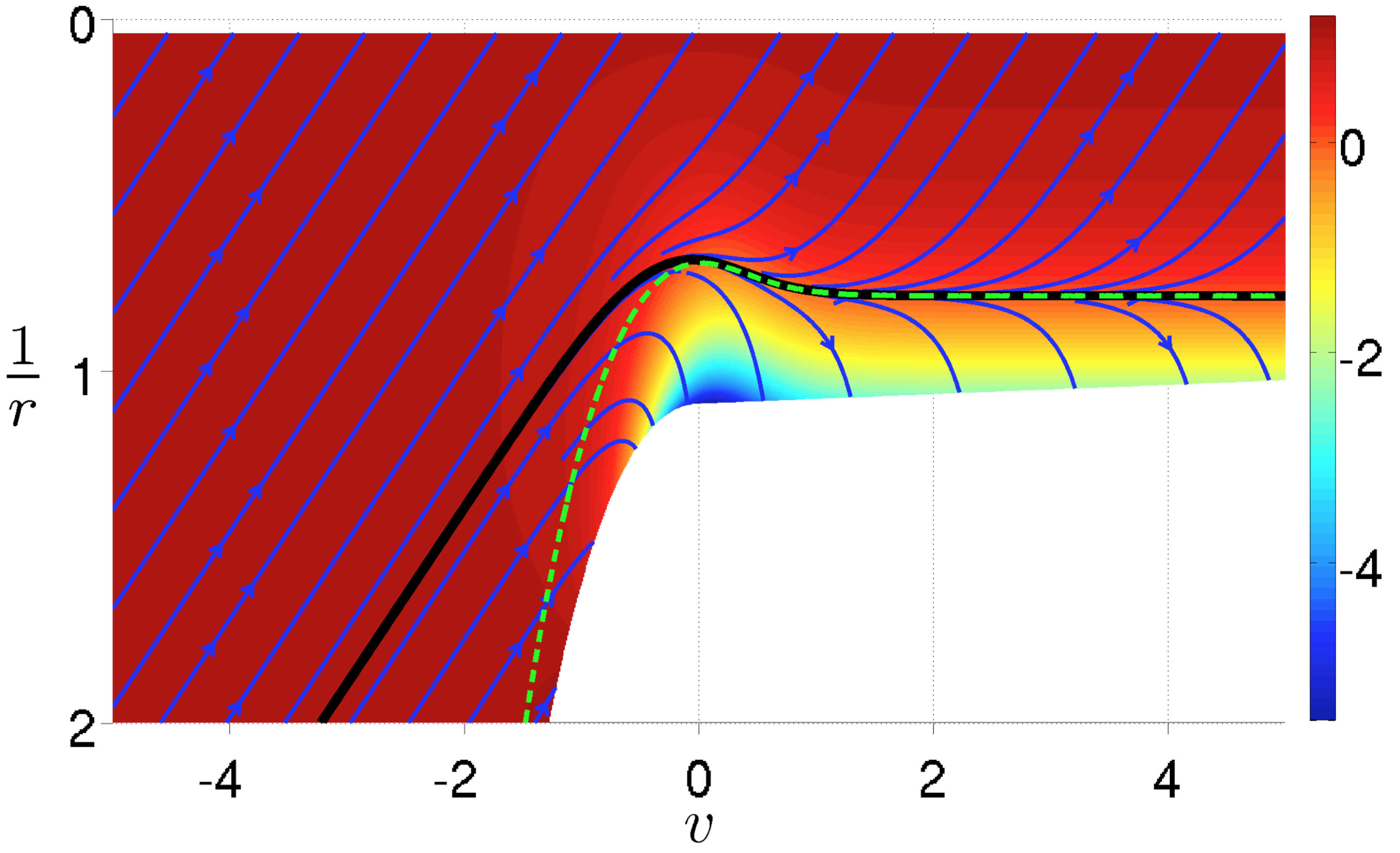}
\end{center}
\caption{
\label{cool_chesler_fig}
Figure from \protect \Ref{Chesler:2008hg}. 
The congruence of outgoing radial null geodesics. 
The surface coloring displays $A(v,r)/r^2$ . The excised region is 
beyond the apparent horizon, which is shown by the dashed 
green line. The geodesic shown as a solid black line is the 
event horizon; it separates geodesics which escape to the 
boundary from those which cannot escape. 
}
\end{figure}

Translation invariance allows the $5d$ metric to be written 
\st
 ds^2 =  A dv^2 + 2 dr dv + \Sigma^2\left(e^{B} d\x_{\perp}^2 + e^{-2B} dx_{\parallel}^2 \right) \, , 
\stp
where  all coefficients $A,\Sigma, B$ are functions of radius $r$ and time $v$.
The  metric coefficient $A(v,r)$ together with  the outgoing radial geodesics
calculated in this geometry are shown in \Fig{cool_chesler_fig} and is reproduced from \cite{Chesler:2008hg}.  
Outgoing 
light like geodesics satisfy  $dr/dv =  A/2$.
We will determine the non-equilibrium string correlators in this  transient geometry.

A salient feature of the outgoing geodesics is their ultimate bifurcation
at the event horizon, $r_o(v)$, which is shown by a thick black line in \Fig{cool_chesler_fig}.  This bifurcation is reminiscent of \Fig{geodesics_markup}. It is convenient
to switch coordinates to a system of coordinates where this black line is 
flat
\begin{align}
    \hr =& r - r_o(v).
\end{align}
Note $r_o(v)$ defines a light like outgoing radial geodesic, so that the metric is 
\st
  ds^2 = - (A- A_{o}) dv^2 + 2 \,d\hr\,  dv + \Sigma^2 \left(e^{B} d\x_{\perp}^2 + e^{-2B} dx_{\parallel} \right) \, , 
\stp
where $A_{o}(v) =  A(r_o(v), v)$.  
In the near horizon geometry, where $r\simeq r_o(v)$ we can approximate
\st
 A(r,  v) - A_{o}( v)  \simeq  
\left. \frac{\partial A(r,v) }{\partial r}  \right|_{r= r_h(v)}   \hr \, , 
\stp
where here and below we will define the stretched horizon at 
$\hr_h \equiv \epsilon$, {\it i.e.} $r_h(v) \equiv r_{o}(v) + \epsilon$.
For future convenience we define the ``effective temperature"%
\footnote
  {
  We note that $T_{\rm eff}$ should not be interpreted as a temperature
  at all times.  In particular, due to the teleological nature of 
  event horizons, $T_{\rm eff}$ is non-zero even before the boundary
  geometry has changed -- \textit{i.e.} when the dual boundary 
  quantum theory is still in the vacuum state.  It is only at late times
  when the black hole starts to thermalize that $T_{\rm eff}$ can be interpreted
  as a temperature.
  }
\st
\kp(v ) \equiv  \frac{1}{2} \left. \frac{\partial A(r,v) }{\partial r}  \right|_{r=r_h(v)}     \label{Teff} \, .
\stp
The action of the string fluctuations in the $x$ direction is the 
same as \Eq{sng}, but the metric coefficients depend on time and radius.
With the goal of determining the horizon effective action it is useful 
to define  a non-equilibrium drag coefficient
\st
 \eta(v) \equiv \frac{\sqrt{\lambda}}{2\pi} g_{xx}(r_h(v), v)\,.
\stp

\subsection{Calculation}

The computational procedure for computing correlators 
in the non-equilibrium case is remarkably similar to the equilibrium case discussed in Sections~\ref{sec2} and \ref{HorizonEFT}.
For both $G_{rr}$ and $G_{ra-ar}$ one can write down a solution to the equations 
of motion in terms of retarded Green functions
convoluted with initial data, as in Eq.~(\ref{formal}).  Furthermore, for initial data
specified suitably far in the past, the relevant initial data for evolution near $v=0$ (when the boundary geometry is changing)
will come from a narrow strip near the horizon.%
\footnote
  {
  To make this more precise, suppose instead of 
  starting off with an initial geometry which was $AdS_5$, the initial geometry consisted of a static black hole
  geometry at temperature $T_{\rm initial} = \delta$.  Assuming initial data is specified at times $v \ll -1/\delta$ 
  in the past, all relevant initial data for future evolution around $v = 0$ will come from an exponentially narrow strip which 
  is exponentially close 
  to the horizon.  Of course, one can always consider the limit $\delta \to 0$ after all calculations are performed.
  }
One can then repeat the analysis of Section~\ref{strategy} and conclude that $G_{rr}$ and $G_{ra-ar}$ are determined by 
horizon correlators $G_{rr}^h$ and $G_{ra-ar}^h$, as in Eqs.~(\ref{convolution}) and (\ref{convolutionspec}).

In the non-equilibrium case $G_{rr}^h$ is given by 
\st
\label{noneqgrrh}
  G_{rr}^h(v_1, v_2) =   \left[  - \eta(v_1) \sqrt{h} h^{\hr \hr}(\hr_1) \partial_{\hr_1}  \right]\left[-\eta(v_2) \sqrt{h} h^{\hr \hr}(\hr_2) \partial_{\hr_2} \right]  g_{rr} (v_1 \hr_1| v_2 \hr_2)
\stp
where $g_{rr}(v_1 \hr_1| v_2 \hr_2)$ is a solution of the homogeneous 
equations, but is confined to the strip,  $0 < \hr \le \epsilon$. 
Furthermore $g_{rr}(v_1 \hr_1| v_2 \hr_2)$ 
should satisfy the boundary conditions  $g_{rr}(v_1,\hr_1=\epsilon|v_2 ,
\hr_2) = g_{rr}(v_1\hr_1|v_2\hr_2=\epsilon) = 0$. 

The equations of motion for $g_{rr}(\hv_1 \hr_1 | \hv_2 \hr_2)$  are
\st
\left[ \pp{\hv_1}  g_{xx} \sqrt{h} h^{\hr v} \pp{\hr_1} + \pp{\hr_1} g_{xx} \sqrt{h} h^{\hr v} \pp{\hv_1 } + \pp{\hr_1} g_{xx}  \sqrt{h} h^{\hr \hr} \pp{\hr_1} \right] g_{rr}(\hv_1 \hr_1 |  \hv_2 \hr_2) = 0\, , 
\stp
where all the metric coefficients depend on $v$ and $r$.
Without approximation, we have
\begin{multline} \label{eomnoneq2}
   \frac{\partial \phantom{\rho}}{\partial \rho_1 } \left[ 2 g_{xx} \sqrt{h} h^{\hr v} \pp{\hv_1 } + 
  \frac{ \partial (g_{xx} \sqrt{h} h^{\hr v})}{\partial \hv_1}  + 
  g_{xx} \sqrt{h} h^{\hr\hr}  \pp{\hr_1} \right] g_{rr}(\hv_1\hr_1 | \hv_2 \hr_2 )    \\
- \left[ \frac{\partial (g_{xx} \sqrt{h} h^{\hr v})  }{\partial \hr_1}   \frac{ \partial  }{\partial \hv_1 }  
+  \frac{ \partial(g_{xx}{\sqrt{h }h^{\hr v}) }}{\partial \hr_1 \partial \hv_1 } \right] g_{rr}(v_1\hr_1| v_2\hr_2)  = 0 \, .
\end{multline}

The above equation of motion contains both first and second derivatives.  However, as we are interested in 
solutions which are irregular near the horizon, all terms with single derivative operators acting on $g_{rr}(v_1\hr_1| v_2\hr_2)$
can be neglected.  This approximation leads to
\st \label{noneqeqm}
   \frac{\partial \phantom{\rho}}{\partial \rho_1 } \left[ 2\eta \sqrt{h} h^{\hr v} \pp{\hv_1 } + 
  \frac{\partial (\eta \sqrt{h} h^{\hr v})}{\partial \hv_1}  + 
  \eta \sqrt{h} h^{\hr\hr}  \pp{\hr_1} \right] g_{rr}(\hv_1 \hr_1 | \hv_2 \hr_2 ) = 0 \, ,   
\stp
and an analogous equation for $v_2$ and $\rho_2$.
Inspecting the above equation, we see that 
$\sqrt{\eta(v_1)\eta(v_2)} g_{rr}(\hv_1 \hr_1 | \hv_2 \hr_2 )$ is annihilated by the operator
$\partial_\rho \left[\partial_v + \frac{1}{2} A \partial_\rho \right]$.  As in Section~\ref{sec:horizonsymcorr}, 
the consequence of this is that any function which is constant on null radial
geodesics satisfies the equations of motion.  Near the horizon this translates to 
$\sqrt{\eta(v_1) \eta(v_2)} g_{rr}(\hv_1 \hr_1 | \hv_2 \hr_2 )$ being 
a function of $v$ and 
\st
\label{routeq}
 \rout(\hr,v) = \hr \, e^{-\int^{v}_{v_o} \kp(v') dv' } \, .
\stp
As in Eq.~(\ref{feqn}), the general solution to Eq.~(\ref{noneqeqm}) which satisfies the requisite boundary conditions at the stretched horizon reads
\begin{align} \label{grrnoneqsol}
\sqrt{\eta(v_1) \eta(v_2) } g_{rr}(v_1\hr_1| v_2 \hr_2) 
  =& f( \rout(\hr_1,v_1), \rout(\hr_2,v_2) )  - f( \rout(\hr_h,v_1), \rout(\hr_2,v_2) )  \nonumber \\
  &- f( \rout(\hr_1,v_1), \rout(\hr_h,v_2) )  + f( \rout(\hr_h,v_1), \rout(\hr_h,v_2) )   \, ,
\end{align}
where as in the equilibrium case, $f(x_1,x_2)$  is determined by the 
initial conditions specified in the distant past.

To determine $f(x_1,x_2)$ we invoke a similar argument used in Section~\ref{sec:horizonsymcorr}.  In particular, assuming 
initial data for $g_{rr}(v_1\rho_1|v_2 \rho_2)$ was specified in the distant past, the only relevant 
initial data for evolution near $v=0$ 
comes from a very narrow strip which is very close to the event horizon.
In this strip the relevant initial data is given by the coincident point limit of the symmetrized correlator as given in Eq.~(\ref{coincidentpts}).
Consequently $f(x_1,x_2)$ must be given by
\begin{equation}
\label{fasymnoneq}
f(x_1, x_2)  =  -\frac{1}{4\pi} \log|x_1 - x_2| \, .
\end{equation}

With $g_{rr}(v_1\rho_1|v_2\rho_2)$ known, \Eq{noneqgrrh} yields the symmetrized horizon correlator 
\begin{align}
\label{GrrH_final}
 G_{rr}^h(\hv_1,\hv_2) =& 
-\frac{\sqrt{\eta(v_1) \eta(v_2)} }{\pi}
\partial_{v_1} \partial_{v_2} \log | 1 - 
 e^{- \int_{v_1}^{v_2} \kp(v') dv' } | \, .    
\end{align}
We will discuss the physical implications of this result in the next section.

The horizon spectral density can be obtained along the same lines
as the horizon symmetrized correlator with the subscript replacement $rr \to ra-ar$ in Eq.~(\ref{noneqgrrh}).
Because $g_{ra-ar}(v_1 \rho_1| v_2 \rho_2)$ satisfies the same equations of motion and boundary conditions
as $g_{rr}(v_1 \rho_1| v_2 \rho_2)$, the general solution for $g_{ra-ar}(v_1 \rho_1| v_2 \rho_2)$ takes the same form as Eq.~(\ref{grrnoneqsol}).
However, as in the equilibrium case discussed in Section~\ref{sec:horizonspectraldensity}, the initial conditions determining $f(x_1,x_2)$
for the spectral density come purely from the equal time canonical commutation relations and yield
\begin{equation}
f(x_1,x_2) =   \frac{1}{4} {\rm sign} (x_1 - x_2) \, .
\end{equation}
With $g_{ra-ar}$ known, the horizon spectral correlator reads
\st
    G_{ra - ar}^h(v_1,v_2) = 2\sqrt{\eta(v_1) \eta(v_2) } \delta'(v_1-v_2) \, . 
\stp

\section{Summary}
\label{summary}

Let us summarize our results.  For definiteness we will describe how to compute
string fluctuations in the non-equilibrium geometry determined by  Chesler and
Yaffe \cite{Chesler:2008hg}.  We emphasize, however,  that similar formulas can
be used for other fields such as gravitons and dilatons.
\Fig{cool_chesler_fig} shows the event horizon $r_{o}(v)$ together with the
associated bifurcating outgoing geodesics of this geometry.   The stretched horizon is located at
$r_{h}(v) = r_o(v) + \epsilon$ 

In contrast to a retarded Green function (which is a response to a source),
a symmetrized correlation function, or a fluctuation,
 is a time dependent correlation that arose
from a definite initial condition.
In the case of Hawking radiation (see \Fig{geodesics_markup}), 
this initial condition is the result of an ultra-violet
vacuum fluctuation (or a symmetrized correlation function) 
which originated close
to the event horizon of the developing black hole. This UV fluctuation skims
the event horizon following outgoing lightlike geodesics until late times.
Then the fluctuation, which is no longer so ultra-violet, leaves the bifurcating horizon  and induces stochastic motion in
the string. We implemented this picture of Hawking radiation directly.

We showed how the Hawking flux out of equilibrium can  be
packaged into an effective action on the stretched horizon which 
can be used to determine the effect of the Hawking radiation on the exterior
dynamics. The horizon action through quadratic order  
is
\begin{align}
 iS_{\rm eff}[x^h] = - \frac{1}{2} \int {\dd v_1}{\dd v_2}\, x_a^h(v_1) &\left[iG_{ra-ar}^{h}(v_1,v_2) \right] x_r^h( v_2)  \nonumber  \\
 &- \frac{1}{2} \int \dd v_1 \dd v_2 \, x_{a}^h(v_1) \left[  G_{rr}^h(v_1,v_2) \right]
x_a^h(v_2) \, ,  
\end{align}
where the horizon correlation functions are the horizon spectral density,
\st
     G_{ra-ar}^h(v_1,v_2) = 2\sqrt{\eta(v_1) \eta(v_2) } \delta'(v_1- v_2)  \, , 
\stp
and the horizon symmetrized correlator, 
\st
 G_{rr}^h(\hv_1,\hv_2) = 
-\frac{\sqrt{\eta(v_1) \eta(v_2)} }{\pi}
\partial_{v_1} \partial_{v_2} \log | 1 - 
 e^{- \int_{v_1}^{v_2} \kp(v') dv' } | \, .    
\stp
Here  $x^h(v)$ is the location of the string on the stretched horizon as
a function of time.
The  coefficient 
$\eta(v) = (\sqrt{\lambda}/2\pi) g_{xx}(v,r_h(v))$ determines
the coupling between the world sheet fluctuations and 
the near horizon geometry, and 
the effective horizon temperature  records 
the Lyapunov exponent of diverging geodesics along the bifurcating
horizon (see \Eq{Teff} and \Eq{routeq}).  More invariantly,  it is related to the extrinsic curvature on the stretched horizon.
In equilibrium, $\eta(v)$ is the drag of the heavy quark,  
and $T_{\rm eff}$  is the Hawking temperature.
The horizon spectral density (which is proportional to $\omega$) is determined by the canonical commutation relations of the
$1+1d$ effective theory which describes the near horizon dynamics. The
horizon symmetrized  correlator is determined by the initial density matrix of the
effective theory far in the past.
When  the horizon action is coupled to the bulk, this action
generates noise on the stretched horizon which induces the random motion of the 
quark in the dual field theory. 
For the stationary black hole the effective action for the string was
determined previously using complicated analytic continuations \cite{Son:2009vu}. 

The importance of these horizon
functions is that they determine the 
spectral density and 
symmetrized correlations
in  the bulk and boundary theories. Indeed, the bulk spectral
density (the commutator) and symmetrized correlator (the anti-commutator)  are determined by propagating their horizon counterparts away from 
the stretched horizon
\st
 iG_{ra-ar}(v_1r_1|v_2r_2) = \int \dd v_1' \dd v_2' \, G_{ra}(v_1r_1|v_1'r_h(v_1')) \, G_{ra}(v_2r_2|v_2'r_h(v_2')) \, iG_{ra-ar}^h(v_1'|v_2') \, ,
\stp
\st
 G_{rr}(v_1r_1|v_2r_2) = \int \dd v_1' \dd v_2' \, G_{ra}(v_1r_1|v_1'r_h(v_1')) \; G_{ra}(v_2r_2|v_2'r_h(v_2'))  \; G_{rr}^h(v_1'|v_2') \, .
\stp
Finally,
these bulk correlation functions can be lifted to the boundary 
to determine the spectral density 
and symmetrized correlator in the field theory (see \app{bndbulk}).
When the fluctuations are thermalized, the two correlation functions satisfy the
fluctuation dissipation theorem
\st
  G_{rr}(\omega,r_1,r_2)  = \left(\frac{1}{2} + n(\omega)  \right) iG_{ra-ar}(\omega,r_1,r_2)  \, .
\stp
Thus we can numerically determine the fluctuations and monitor their
approach to equilibrium using the formalism of this work. This 
numerical calculation will be presented in future work.


Even without a complete numerical computation, some preliminary remarks
can be made about equilibration in $AdS_5$.  
The Wigner transforms
 of $G_{rr}^h(v_1,v_2)$ and $G_{ra-ar}^h(v_1,v_2)$
obey the fluctuation
dissipation theorem at {\it  high} frequency.%
\footnote{The Wigner transform is the Fourier transform with respect to the difference $v_1 - v_2$, as a function of the average, $\bar{v} = (v_1 + v_2)/2$.} 
 Specifically, for a typical non-equilibrium
time scale $\tau$,  we have
\st
    G_{rr}^h(\bar{v}, \omega) \simeq  \left(\frac{1}{2} + n(\omega) \right)  iG_{ra-ar}^h(\bar v, \omega) \,  + \, O\left(\frac{1}{\tau^2 \omega^2} \right) \, ,
\stp
where $\bar{v} = (v_1 + v_2)/2$.
Thus the string is born into equilibrium at high frequency, and eventually frequencies of order the temperature and below equilibrate.  This conclusion seems squarely aligned with the results of \Ref{Balasubramanian:2010ce} which 
was limited to operators of high conformal dimension. However, 
it must be emphasized that the map between the stretched horizon and the boundary 
is non-trivial, especially for five-dimensional 
fields where  the 3-momentum  influences the coupling between
the field and the near horizon geometry.

One popular picture of Hawking radiation is based on quantum tunneling (see,
for instance, \cite{Parikh:1999mf}).  This picture relates the thermal factor
in the emission rate to the change in the black hole entropy, which appears by
way of the Euclidean action.  It would be interesting to make contact with this
picture using the effective action formalism. This would help make the
universality of the result fully manifest.  It would also be interesting to see
if the simple non-equilibrium effective action presented in this paper finds a
simple origin in the tunneling picture.

The current derivation of Hawking radiation and correlation functions 
is similar  to the  2PI formalism of non-equilibrium field theory \cite{Berges:2004yj}, 
and we hope this will make black hole physics accessible to a wider 
audience.
The derivation uses the unstable 
nature of the bifurcating event horizon 
to expand  ultraviolet vacuum fluctuations. This exponential sensitivity to
initial conditions has been  called the transplanckian  problem
\cite{Jacobson:2003vx}, and is characteristic of classically chaotic systems
\cite{PhysRevE.51.28}. By exploiting this analogy, 
we hope that new insight can be found into the transplanckian problem 
and the Bekenstein entropy. Understanding the Bekenstein-Hawking entropy 
will require coupling the particle emission to 
the background metric,  leading to a dynamical competition
between quantum particles and the classical background field. 
The particle-field problem has been extensively studied 
in thermal field theory and the Color Glass Condensate \cite{Mueller:2002gd,Dusling:2010rm}.  We
hope to pursue these connections in future work. \\
{}\\
\noindent{ \bf Comparison with recent literature: } \\

Recently several papers appeared which addressed aspects of this paper 
as this work was being finalized \cite{Ebrahim:2010ra}.
First, a paper by Headrik and  Ebrahim \cite{Ebrahim:2010ra} used the equations of motion to solve
for the symmetrized correlation function in $AdS_3$-Vaidya spacetimes.  
Headrik and Ebrahim 
reported on the ``instantaneous thermalization" of $AdS_3$.
Second, a paper  by  Balasubramanian {\it et al} \cite{Balasubramanian:2010ce} 
computed the thermalization of operators 
with high conformal dimension by studying geodesics. Although the current paper
is not limited to such operators, the basic conclusion that
the field theory thermalizes first at high frequency is 
consistent with our conclusion about horizon Wigner transforms. 
However, as emphasized above the map between the stretched horizon
and the boundary involves physically important and non-trivial outgoing propagators. \\
{}\\
\noindent{ \bf Acknowledgments:} \\
{}\\
We would like to thank John McGreevy for useful discussions.
The work of SCH is supported by NSF grant PHY-0969448. PMC is supported by a Pappalardo Fellowship at MIT.
DT is supported in part by the Sloan Foundation and 
by the Department of Energy through the Outstanding Junior Investigator 
program, DE-FG-02-08ER4154.

\appendix

\section{Symmetrized correlations  and the Keldysh formalism -- a lesson from the harmonic oscillator }
\label{shoapp}

It is instructive in many respects to compute the symmetrized correlation
function of the harmonic oscillator using the Keldysh formalism. 
The action of the oscillator  is 
\st
  iS_1 - iS_2 =  i\int_{t_o} \dd t \left[ m \dot x_r \dot x_a - m \omega_o^2 x_r x_a \right] \, ,
\stp
and is   similar to the string action written in  \Eq{actions1s2}.
In both systems we see that   
there are no $rr$ type propagators in the action itself.
This is because 
symmetrized correlations are  the result of a correlation built into the initial state wave functions, {\it i.e.} the initial density matrix.

In symmetrized type  correlation functions the
density matrix at an initial time $t_o$ 
correlates the initial conditions for subsequent evolution. 
The density matrix for the harmonic oscillator in the ground state
is $\Psi_{o}(x_1) \Psi_o^*(x_2)$, and the symmetrized correlator  is
\st
 G_{rr}(t,\bar{t})  =  
\int \dd x_r^o \dd x_{a}^o \; 
\Psi_{o}(x_r^o + x_a^o/2)  
\Psi_o^{*}(x_r^o - x_a^o/2) 
\int_{x_r^o, x_{a}^o}  \Dt x_r \Dt x_a e^{iS_1 - iS_2}  
 x_r(t) x_r(\bar t) \, .
\stp 
This  is simplified by (i) introducing the Wigner transform,
\[
  W(x_r^o,p^o) = \int \dd x_a^o \, e^{-ip^o x_a^o} \,  
\Psi_{o}(x_r^o + x_a^o/2)  
\Psi_o^{*}(x_r^o - x_a^o/2)   \, , 
\]
(ii) integrating by parts in the action (with the boundary condition $x_a \rightarrow 0$ for $t\rightarrow +\infty$), 
\[
 iS_1 - iS_2 = -i m x_a^o \dot x_r(t_o) - i \int_{t_o} \dd t \, x_a \left[ m\ddot x  + m\omega_o^2 x \right] \, , 
\] 
and finally (iii) integrating over 
all $x_a$, yielding
\st
G_{rr}(t, \bar t)
= \int \frac{\dd x_r^o  \dd p^o}{2\pi } W(x^o_r, p^o) \,  \delta( p^o - m \dot x_r(t_o))  \int_{x_r(t_o) = x_r^o}\Dt x_r \, \delta_t \left[ M \ddot x_r + \m\omega_o x_r \right]\, x_r(t) x_r(\bar t) \, .
\stp
Here $\delta_t[f(t)]$ denotes the functional delta function, $\prod_{t}\, \delta (f(t))$.
The meaning of this path integral is that $G_{rr}(t,\bar t)$ is found
by solving the equation of motion for a specified initial condition,
\st
  x_r (t) = x_{r}^{o} \cos(\omega_o (t-t_o)) + \frac{p^o}{m \omega_o} \sin(\omega_o (t-t_o)) \, ,   \label{eq_cauchy}
\stp
and then averaging the square of this  solution over 
the initial conditions specified by the Wigner transform.
Performing this average for the ground state wave function of the 
oscillator reproduces the familiar result
\st
  G_{rr}(t,\bar t) = \frac{1}{2} \llangle 0 | \left\{ \hat{x}(t), \hat{x}(\bar t) \right\} | 0 \rrangle 
 =  \frac{1}{2 m\omega_o} \cos(\omega_o(t-\bar t) )  \, .
\stp
The lesson from this analysis is that symmetrized correlation functions
invariably arise from  correlations in the initial density matrix 
which are propagated forward by the equations  of motion. 
This dependence on the initial density matrix should be contrasted with 
retarded propagators which are independent of the wave function
of the simple harmonic oscillator, {\it i.e. }
\st
    \theta(t-\bar t) \left[ \hat x(t)  , \hat x(\bar t) \right] =  \frac{-i \theta(t -\bar t)}{m \omega_o} \sin(\omega_o(t-\bar t) ) \, ,
\stp
is a  pure number. 

\section{The Green Function Composition Rule}
\label{wronskianapp}

In this appendix we detail the Green function composition rule stated in \Eq{composition}.
In this section $\one,\two,\three$ denote the space-time points, {\it e.g.} $\one =  (v_1, r_1)$.

Suppose $G_{ra}(\one|\two)$ and $\hat{G}_{ra}(\one|\three)$ are retarded Green functions.
Then the Wronskian of  the two Green functions is
\st
\label{Wronsk1}
  W^{\mu}(\two) =   \frac{\sqrt{\lambda}}{2 \pi } g_{xx} \sqrt{h} h^{\mu\nu}(\two) \left[ G_{ra}(\one|\two) \frac{\ddiff}{\partial 2^{\nu}}  \hat{G}_{ra}(\two|\three) \right] \, , 
\stp
where  $\ddiff = \overrightarrow{\partial}-\overleftarrow{\partial}$ and is not intended to act outside of the square braces.
Then, a short exercise  shows that the divergence is 
\st
 \partial_{\mu}W^{\mu}(\two)
= G_{ra}(\one|\three) \delta (\two,\three)  - \hat G_{ra}(\three|\one) \delta (\two , \one) \,  .
\stp
Assuming that $\three$ is inside the strip and $\one$ is outside the strip (see \Fig{geo2_markup}) we can integrate over the strip to obtain the retarded Green function:
\begin{align}
\label{divergence1}
G_{ra}(\one|\three) =  \int_{r < r_h} \frac{\partial}{\partial 2^{\mu}} W^{\mu}(\two) 
 =& \int_2 \dd\Sigma_{\mu} \,  \frac{\sqrt{\lambda}}{2\pi} g_{xx} \sqrt{h} h^{\mu\nu}(\two)  \left[ G_{ra}(\one|\two) \frac{\ddiff}{\partial 2^{\nu}}  \hat G_{ra}(\two|\three) \right] \,  ,
\end{align}
where $\dd\Sigma_{\mu}$ is a surface surrounding the 
strip  with outward directed normal, and the integration is over space-time point $2$.
We next use the near horizon approximation for the leading factors (\Eq{drag_coeff}),   
and neglect all surface
terms  except the integral over the stretched horizon.  These surface integrals vanish because one of the  Green functions vanishes. For instance,
on the past surface (where $v=-\infty$)   $\hat G_{ra}(\two|\three)$ must vanish
since it represents the causal response at point $\two$ (past infinity) to a
source at point $\three$. 
Since we have not specified the boundary conditions on the retarded Green function $\hat{G}_{ra}$, we are free to specify reflective Dirichlet boundary conditions on the stretched horizon, {\it i.e.} take $\hat{G}_{ra}(\two|\three)  ={g}_{ra}(\two|\three)$, as defined in the text. This specification  does not interfere with the relevant initial data extremely close to the real event horizon. 
With this choice \Eq{divergence1} results in \Eq{composition} given 
in the text.

%

\section{From bulk to bulk correlators, to boundary to bulk correlators,
to CFT correlators }
\label{bndbulk}
\def\bdy{{\rm{bnd\,to\,bulk}}}  

Given a retarded bulk to bulk correlator $G_{ra}(\one|\two)$ 
in the 
gravitational theory,  we will show that  the retarded correlator
in the CFT  is
\st
  G_{ra}^{CFT}(t_1|t_2) = \lim_{r_1 \rightarrow  {\rm bnd}} \; \lim_{r_2 \rightarrow
{\rm bnd}} \left[-\frac{\sqrt{\lambda}}{2\pi}  g_{xx} \sqrt{h} h^{rr}(\one)
\partial_{r_1} \right] \left[-\frac{\sqrt{\lambda}}{2\pi } g_{xx} \sqrt{h}
h^{rr}(\two) \partial_{r_2} \right] \,  G_{ra}(\one|\two) \, .   \label{CFTgra}
\stp
In holography, the CFT correlation function is usually determined by setting the boundary conditions of five dimensional fields as $r\rightarrow\infty$.
We claim that  \Eq{CFTgra} is an equivalent prescription. Indeed,  \Eq{CFTgra}
arises if the path integral discussion of the 
horizon effective action in \Sect{horizon_effective} is applied to the boundary.  
Specifically, instead of integrating out the strip to find an effective 
action of the stretched horizon, one integrates out the entire 
bulk to find an effective action of the CFT.
Put differently, from the viewpoint of the bulk theory, any small change in a boundary condition
must be equivalent to inserting a suitable local operator close to that boundary, and for a Dirichlet condition that operator turns out to be a derivative of the field as in \Eq{CFTgra}
(see, for instance, \cite{symanzik}).

To make direct contact with the pioneering work of Policastro, Son, and Starinets \cite{Policastro:2002se}, we define
the boundary to bulk correlator $f(t_1r_1|t_2)$ as a retarded solution  which satisfies
\st
\label{bboundary}
  \lim_{r_1 \rightarrow {\rm bnd}} f(t_1r_1|t_2)  = \delta(t_1 - t_2) \, .
\stp
For static geometries in Fourier space this is usually called $f_{\omega}(r_1)$,
and the retarded correlator in the CFT is usually  
\st
  G_{ra}^{CFT} (\omega) = + \frac{\sqrt{\lambda}}{2\pi} g_{xx} \sqrt{h} h^{rr} \partial_r f_{\omega}(r_1) \, .
\stp
Now we claim that  the boundary to bulk propagator is simply related to the bulk to bulk propagator via
\st
\label{frelation}
  f(t_1r_1|t_2) = \lim_{r_2 \rightarrow {\rm bnd} }  \frac{\sqrt{\lambda}}{2\pi}\sqrt{h}h^{rr}(r_2) \partial_{r_2}  G_{ra}(t_1r_1|t_2r_2) \, .
\stp
To show this  we take $r_1$ fixed and large, and  we integrate the equations of
motions of the retarded Green function (\Eq{eomgreen}) with respect to the second
argument over the pill-box shown below.  The radius of the lower surface is small compared to
$r_1$ (but still  large), while the radius of the upper surface is large compared to $r_1$.
\st
 \includegraphics[width=2.0in]{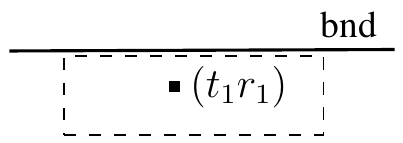}
\stp
This  yields
\begin{multline}
\label{limits}
  \lim_{r_1 \rightarrow {\rm bnd} } \lim_{r_2 \rightarrow {\rm bnd} }  
 \frac{\sqrt{\lambda}}{2\pi} g_{xx} \sqrt{h} h^{rr}(r_2) 
\frac{\partial}{\partial r_2} G(r_1 t_1|r_2 t_2)  \\
 - \lim_{r_2\rightarrow {\rm bnd} } \lim_{r_1\rightarrow {\rm bnd} }
\frac{\sqrt{\lambda}}{2\pi} g_{xx} \sqrt{h} h^{rr}(r_2)
\frac{\partial}{\partial r_2 } G(t_1r_1|t_2r_2) = \delta(t_1 - t_2) \, .
\end{multline}
The second term vanishes since $G_{ra}(t_1r_1|t_2 r_2)$ obeys 
Dirichlet boundary conditions.  Thus, the first term in \Eq{limits} 
obeys the same  boundary conditions as the bulk to boundary propagator 
$f(t_1r_2|t_2)$, \Eq{bboundary}. Since both functions
are retarded and obey the same equations of motion and boundary conditions, they are  
the same  and \Eq{frelation} holds.
By extension, \Eq{CFTgra} for the CFT correlators from bulk to bulk correlators is equivalent to the usual prescription.



\bibliography{njp}

\end{document}